\documentclass[aps,prc,10pt,showpacs,showkeys,twocolumn,superscriptaddress,groupedaddress]{revtex4-1}
\usepackage[latin1]{inputenc}
\usepackage[english]{babel}
\usepackage{amsmath}
\usepackage{amssymb}
\usepackage{amscd}
\usepackage{eucal}
\usepackage{color}
\usepackage{bm}
\usepackage{braket}
\usepackage{dsfont}
\usepackage{float}
\usepackage{graphicx}
\usepackage[makeroom]{cancel}
\usepackage{bbm}
\usepackage[capitalize]{cleveref}
\bibstyle{asprev}

\newcommand{\nmax}{N_{\rm max}}
\newcommand{\hb}{\hbar\Omega}

\begin{document}

\title{Nuclear kinetic density from {\it ab initio} theory}

\author{Michael Gennari}
\email{mgennari5216@gmail.com}
\affiliation{University of Waterloo, 200 University Avenue West, Waterloo, Ontario N2L 3G1, Canada \\
TRIUMF, 4004 Wesbrook Mall, Vancouver, British Columbia V6T 2A3, Canada}

\author{Petr Navr\'atil}
\email{navratil@triumf.ca}
\affiliation{TRIUMF, 4004 Wesbrook Mall, Vancouver, British Columbia V6T 2A3, Canada}

\date{\today}

\begin{abstract}

\noindent
{\bf Background:} The nuclear kinetic density is one of many fundamental, non-observable quantities in density functional theory (DFT) dependent on the nonlocal nuclear density. Often, approximations may be made when computing the density that may result in spurious contributions in other DFT quantities. With the ability to compute the nonlocal nuclear density from \textit{ab initio} wave functions, it is now possible to estimate effects of such spurious contributions.

\noindent
{\bf Purpose:} We derive the kinetic density using {\it ab initio} nonlocal scalar one-body nuclear densities computed within the no-core shell model (NCSM) approach, utilizing two- and three-nucleon chiral interactions as the sole input. The ability to compute translationally invariant nonlocal densities allows us to gauge the impact of the spurious center-of-mass (COM) contributions in DFT quantities, such as the kinetic density, and provide \textit{ab initio} insight into refining energy density functionals.

\noindent
{\bf Methods:} The nonlocal nuclear densities are derived from the NCSM one-body densities calculated in second quantization. We present a review of COM contaminated and translationally invariant nuclear densities. We then derive an analytic expression for the kinetic density using these nonlocal densities, producing an \textit{ab initio} kinetic density.

\noindent
{\bf Results:} The ground state nonlocal densities of \textsuperscript{4,6,8}He, \textsuperscript{12}C, and  \textsuperscript{16}O are used to compute the kinetic densities of the aforementioned nuclei. The impact of COM removal techniques in the density are discussed and compared to a procedure applied in DFT. The results of this work can be extended to other fundamental quantities in DFT.

\noindent
{\bf Conclusions:} The use of a general nonlocal density allows for the calculation of fundamental quantities taken as input in theories such as DFT. This allows benchmarking COM removal procedures and provides a bridge for comparison between \textit{ab initio} and DFT many-body techniques.

\end{abstract}

\pacs{}

\maketitle

%%%% introduction
\section{Introduction}

In this paper, we derive an analytic expression for the kinetic density of a nuclear system from the nonlocal scalar one-body nuclear densities calculated in the no-core shell model (NCSM) ~\cite{barrett2013ab} approach.  In particular, we use the method introduced in Ref.~\cite{gennari2018microscopic} to construct the microscopic nonlocal one-body density. The {\it ab initio} NCSM is a rigorous many-body technique which treats all $A$ nucleons as active degrees of freedom and takes realistic two- and three-nucleon forces as the sole input. The model is suited for the description of light nuclei ($ A \lesssim 16 $) as it is able to account for many-nucleon correlations producing high-quality wave functions.

With the NCSM nuclear densities it is possible to compute quantities fundamental to density functional theory (DFT), such as the kinetic density, using \textit{ab initio} wave functions. In systems such as $^{16}$O, this can allow for direct comparison between COM removal procedures in different many-body techniques, such as DFT. DFT is a many-body method for calculating nuclear properties across the nuclear chart, which has been practiced in nuclear physics for approximately 40 years ~\cite{duguet2014nuclear,vautherin1970hartree,vautherin1972hartree,vautherin1973hartree,parr1994density,gross2013density,bhattacharyya2005kinetic}. It involves the minimization of an energy functional with respect to several system densities (kinetic, spin, isospin, etc.).

DFT has had great success and has made significant progress in the description of medium- to heavy-mass nuclei ~\cite{hofmann1998hartree,erler2012limits,kortelainen2010nuclear,kortelainen2012nuclear,kortelainen2014nuclear}. In these types of systems, contributions from the center-of-mass (COM) and the nonlocality in the density are reduced in effect. However, if DFT is to extend its reach to more systems, the size of these two effects must be reviewed to ensure they are under control.

Recently, there has been a significant effort to construct a bridge between DFT and \textit{ab initio} approaches to the nuclear many-body problem~\cite{microscopicEDF2018,lacroix2017bare,grasso2018effective,bonnard2018effective}, with the ultimate goal of reducing the phenomenological nature of DFT and creating direct connections to the underlying quantum chromodyanimcs (QCD). Motivated by this effort, we attempt to contribute to the construction of such a bridge by computing a fundamental input of energy density functionals, the nuclear kinetic density, from \textit{ab initio} many-body wave functions with chiral potentials as the sole input. In this work, we study $^{16}$O and other light nuclei using the NCSM many-body method to explore the effects of nonlocality and COM contamination in the nuclear kinetic density.

The paper is organized as follows: in Sec.~\ref{sec_theory}, divided into three subsections, we focus on the theoretical construction of the kinetic density. The NCSM formalism is discussed in Sec.~\ref{sec_ncsm}, the derivation of the nonlocal translationally invariant density is reviewed in Sec.~\ref{sec_nonlocaldensity}, and finally the construction of the kinetic density from the NCSM nonlocal densities in Sec.~\ref{sec_kineticdensity}. In Sec.~\ref{sec_results} we discuss the nonlocal density results, we present results for the kinetic density, and we compare the contributions of the spurious center of mass (COM) and effects of the removal. In Sec.~\ref{sec_conclusions} we draw our conclusions. In the appendix, Sec.~\ref{sec_appendix}, we present the derivations necessary to this work.

%%%% theoretical framework
\section{Theoretical framework}\label{sec_theory}

%%%$ ncsm theory
\subsection{NCSM}
\label{sec_ncsm}

Evaluating the kinetic density in the most consistent manner from \textit{ab initio} wave functions requires knowledge of the translationally invariant nonlocal density. The $A$-nucleon eigenstates required to calculate the nonlocal one-body density are computed according to the {\it ab initio} NCSM approach~\cite{barrett2013ab}. Within this many-body method, nuclei are considered to be systems of $A$ nonrelativistic point-like nucleons interacting via high-quality, realistic two- and three-body inter-nucleon interactions. Each individual nucleon is treated as an active degree of freedom and the translational invariance of observables, the angular momentum, and the parity of the nucleus under consideration are conserved. The many-body wave function is expanded over a basis of antisymmetric $A$-nucleon harmonic oscillator (HO) states. The basis contains up to $\nmax$ HO excitations above the lowest possible Pauli configuration. The basis is characterized by an additional parameter $\Omega$, the frequency of the HO well. Additional information on the many-body method and the truncation scheme can be found in Ref. \cite{barrett2013ab}.

NCSM wave functions are computed through the diagonalization of the translationally invariant nuclear Hamiltonian, which includes both two- and three-body (NN+3N) forces:
\begin{equation}\label{NCSM_eq}
\hat{H} \ket{A \lambda J^\pi T} = E_\lambda ^{J^\pi T} \ket{A \lambda J^\pi T} \; ,
\end{equation}
with $\lambda$ distinguishing eigenstates with identical $J^\pi T$.

In general, we can accelerate convergence of the HO expansion by applying a similarity renormalization group (SRG) transformation on the $NN$ and $3N$ interactions~\cite{wegner1994flow,bogner2007similarity,roth2008unitary,bogner2010low,jurgenson2009evolution}. While large gains in convergence can be achieved by performing the SRG evolution, it induces higher body terms and can introduce a further dependence on the momentum-decoupling scale $\lambda_{\mathrm{SRG}}$ if the unitarity of the SRG transformation is violated. 

In the present work, we used the NN chiral potential at N\textsuperscript{4}LO with a cutoff $\Lambda = 500$ MeV, recently developed by Entem, Machleidt, and Nosyk~\cite{entem2015peripheral,entem2017high}. This interaction will be denoted as NN-N\textsuperscript{4}LO(500). We also included the three-nucleon potential at the next-to next-to leading order (N\textsuperscript{2}LO) with simultaneous local~\cite{navratil2007local} and nonlocal regularization, a more complete description of which can be found in Ref. \cite{gennari2018microscopic}. The 3N component will be denoted as 3Nlnl, making the notation for the total interaction utilized for the densities NN-N\textsuperscript{4}LO(500)+3Nlnl.

In calculations with the NN-N\textsuperscript{4}LO(500)+3Nlnl interaction, we have worked with $\hb=20$~MeV and $\lambda_{\mathrm{SRG}}$ ranges of 1.6 to 2.0 fm$^{-1}$, both previously determined to be optimal and suitable for the given $s$- and $p$-shell nuclei under consideration \cite{gennari2018microscopic}. Note that the $\nmax=8$ calculations for $^{12}$C and $^{16}$O were obtained using importance-truncated NCSM basis~\cite{roth2007ab,roth2009importance}.

%%%% nonlocal density theory
\subsection{Nonlocal density}\label{sec_nonlocaldensity}

In the current work we use the generalized COM removal method of Ref.~\cite{gennari2018microscopic}, which extended the results of local densities in Ref.~\cite{navratil2004translationally} to generate nonlocal one-body density matrices. We note the differences between this approach and the alternate approaches of Ref.~\cite{burrows2018ab,giraud2008labframe} reside in the method of COM removal, which in this and previously cited works is done directly in coordinate space, while in Ref.~\cite{burrows2018ab,giraud2008labframe} the COM removal from nonlocal and local densities is performed in momentum space.

In coordinate representation, the nonlocal form of the nuclear density operator is defined as
\begin{equation}\label{nonlocdens}
\rho_{op}(\vec{r},\vec{r}\,')=\sum_{i=1}^{A} \big( \, \vert \vec{r} \rangle \langle \vec{r}\,' \vert \, \big)^{i}=\sum_{i=1}^{A} \delta(\vec{r}-\vec{r}_i) \delta(\vec{r}\,'-\vec{r}_i') \, .
\end{equation}
The matrix element of this operator between a general initial and final state obtained in the Cartesian coordinate single-particle Slater determinant (SD) basis is written as
\begin{equation}\label{eqn:wiCOMnlocdens}
\begin{split}
&{}_{SD}\langle A \lambda_j J_j M_j\, \vert \rho_{op}(\vec{r},\vec{r}\,') \vert \, A \lambda_i J_i M_i \rangle_{SD} \\
& = \sum \frac{1}{\hat{J}_f} ( J_i M_i K k \vert J_f M_f )\,\bigg(Y_{l_1}^*(\hat{r})\,Y_{l_2}^*(\hat{r}\,')\bigg)_k^{(K)} \\
&\quad \times R_{n_1,l_1}\big(\vert \vec{r}\, \vert \big) R_{n_2,l_2}\big(\vert \vec{r}\,'\vert \big) \\
&\quad \times (-1)^{l_1+l_2+K+j_2+\frac{1}{2}} \, \hat{j}_1 \hat{j}_2 \hat{K} \left \{
  \begin{tabular}{ccc}
  $j_2$ & $l_2$ & $\frac{1}{2}$ \\
  $l_1$ & $j_1$ & $K$ \\
  \end{tabular}
\right \} \\
&\quad \times \frac{(-1)}{\hat{K}} {}_{SD} \langle A \lambda_f J_f \vert \vert \,(a_{n_1,l_1,j_1}^{\dagger}\,\tilde{a}_{n_2,l_2,j_2})^{(K)}\,\vert \vert A \lambda_i J_i \rangle_{SD} \,. \\
\end{split}
\end{equation}
We suppress the isospin and parity quantum numbers for simplicity. In Eq.~(\ref{eqn:wiCOMnlocdens}), the NCSM eigenstates (\ref{NCSM_eq}) have the subscripts $SD$ denoting that we used Slater determinant HO basis that include COM degrees of freedom as opposed to the translationally invariant Jacobi coordinate HO basis~\cite{navratil2000few}. Further, $\hat{\eta}=\sqrt{2\eta+1}$ and $R_{n,l}(\vert \vec{r} \vert)$ is the radial HO wave function with the oscillator length parameter $b=\sqrt{\frac{\hbar}{m\Omega}}$, where $m$ is the nucleon mass. The one-body density matrix elements are introduced in second-quantization, ${}_{SD} \langle A \lambda_f J_f \vert \vert \,(a_{n_1,l_1,j_1}^{\dagger}\,\tilde{a}_{n_2,l_2,j_2})^{(K)}\,\vert \vert A \lambda_i J_i \rangle_{SD}$. Both $\vec{r}$ and $\vec{r}\,'$ are measured from the center of the HO potential well. As a result of this construction, the density contains a spurious COM component.

We require the removal of the COM component from the nonlocal density if we are to compute a consistent kinetic density. This is enabled by the factorization of the Slater determinant and Jacobi eigenstates,
\begin{equation}\label{eq:sd_jac_fac}
\begin{split}
&\langle \vec{r}_1 \dots \vec{r}_A \vec{\sigma}_1 \dots \vec{\sigma}_A \vec{\tau}_1 \dots \vec{\tau}_A \vert A \lambda J M \rangle_{SD}= \\
&\qquad \langle \vec{\xi}_1 \dots \vec{\xi}_{A-1} \vec{\sigma}_1 \dots \vec{\sigma}_A \vec{\tau}_1 \dots \vec{\tau}_A \vert A \lambda JM \rangle \phi_{000}(\vec{\xi}_0) \, , \\
\end{split}
\end{equation}
with the ground state COM component, labeled in Eq.~(\ref{eq:sd_jac_fac}) as $\phi_{000}(\vec{\xi}_0)$. This is given as the $N=0$ HO state with $\vec{\xi}_0$ proportional to the $A$-nucleon COM coordinate. The matrix element of the translationally invariant operator as given in Ref.~\cite{gennari2018microscopic}, $\rho_{op}^{trinv}(\vec{r}-\vec{R}, \vec{r}^{\, \prime}-\vec{R})$, between general initial and final states is then given by (compare to Eq.~(\ref{eqn:wiCOMnlocdens}))
\begin{equation}\label{eqn:trinvnlocdens}
\begin{split}
&\langle A \lambda_j J_j M_j\, \vert \rho_{op}^{trinv}(\vec{r}-\vec{R},\vec{r}\,'-\vec{R}) \vert \, A \lambda_i J_i M_i \rangle \\
& = \Big(\frac{A}{A-1}\Big)^{\frac{3}{2}}\sum \frac{1}{\hat{J}_f} ( J_i M_i K k \vert J_f M_f ) \\
&\quad \times \big(M^K\big)_{nln'l',n_1l_1n_2l_2}^{-1} \, \bigg(Y_l^*(\widehat{\vec{r}-\vec{R}})\,Y_{l'}^*(\widehat{\vec{r}\,'-\vec{R}})\bigg)_k^{(K)} \\
&\quad \times R_{n,l}\Big(\sqrt{\frac{A}{A-1}} \vert \vec{r}-\vec{R} \vert \Big) R_{n',l'}\Big(\sqrt{\frac{A}{A-1}} \vert \vec{r}\,'-\vec{R} \vert \Big) \\ 
&\quad \times (-1)^{l_1+l_2+K+j_2-\frac{1}{2}} \, \hat{j}_1 \hat{j}_2 \left \{
  \begin{tabular}{ccc}
  $j_1$ & $j_2$ & $K$ \\
  $l_2$ & $l_1$ & $\frac{1}{2}$ \\
  \end{tabular}
\right \} \\
&\quad \times {}_{SD} \langle A \lambda_f J_f \vert \vert \,(a_{n_1,l_1,j_1}^{\dagger}\,\tilde{a}_{n_2,l_2,j_2})^{(K)}\,\vert \vert A \lambda_i J_i \rangle_{SD} \\
\end{split}
\end{equation}
where
\begin{equation}
\label{Mk}
\begin{split}
&\big( M^K \big)_{nln'l',n_1l_1n_2l_2} \\
&= \sum_{N_1,L_1} (-1)^{l+l'+K+L_1} \left \{
  \begin{tabular}{ccc}
  $l_1$ & $L_1$ & $l$ \\
  $l'$ & $K$ & $l_2$ \\
  \end{tabular}
\right \} \hat{l} \hat{l'} \\
&\quad \times \langle nl00l \vert N_1 L_1 n_1 l_1 l \rangle_{\frac{1}{A-1}} \langle n'l'00l' \vert N_1 L_1 n_2 l_2 l' \rangle_{\frac{1}{A-1}} \, \, .
\end{split}
\end{equation}
In Eq.~(\ref{eqn:trinvnlocdens}), the $R_{n,l}\Big(\sqrt{\frac{A}{A-1}} \vert \vec{r}-\vec{R} \vert \Big)$ is the radial harmonic oscillator wave function in terms of a relative
Jacobi coordinate, $\vec{\xi}=-\sqrt{\frac{A}{A-1}}(\vec{r}-\vec{R})$. The $\big( M^K \big)_{nln'l',n_1l_1n_2l_2}$ matrix (\ref{Mk}) introduced in Ref.~\cite{navratil2004translationally} includes generalized harmonic oscillator brackets of the form $\langle nl00l \vert N_1 L_1 n_1 l_1 l \rangle_{d}$ corresponding to a two particle system with a mass ratio of $d$, as outlined in Ref.~\cite{trlifaj1972simple}.

The nonlocal density expressions presented here can be related to the local densities in Ref.~\cite{navratil2004translationally} by restricting the coordinates such that $\vec{r} = \vec{r}\,'$, or
\begin{equation}
\rho(\vec{r}) = \rho(\vec{r},\vec{r}\,') {\vert}_{\vec{r} = \vec{r}\,'} = \rho(\vec{r},\vec{r})  \, .
\end{equation}
The normalization of the nonlocal density is consistent with Ref.~\cite{navratil2004translationally} such that the integral of the local form
\begin{equation}\label{eq:dens_norm}
\int d\vec{r} \, \langle A \lambda J M \vert \rho_{op}(\vec{r}, \vec{r}) \vert A \lambda J M \rangle = A \,
\end{equation}
returns the number of nucleons for both (\ref{eqn:wiCOMnlocdens}) and (\ref{eqn:trinvnlocdens}).

Finally, make note that the proton and neutron densities are obtained separately by introducing ($\frac{1}{2}\pm t_{zi}$) factors, respectively, in Eq. (\ref{nonlocdens}). This results in the inclusion of a proton or neutron index in the creation and anihilation operators, as the COM operators commute with isospin operators. The normalization (\ref{eq:dens_norm}) then becomes $Z$ or $N$ for the proton and neutron density respectively.

%%%% kinetic density theory
\subsection{Kinetic density}\label{sec_kineticdensity}

In DFT, the kinetic density is just one of several system densities which contribute to the local energy density $\mathcal{H}(\vec{r})$. The kinetic density is not itself an observable, however when combined with the potential interaction terms, the resultant local energy density $\mathcal{H}$ is an observable from which nuclear properties can be computed ~\cite{borycki2006pairing}. The kinetic term in $\mathcal{H}(\vec{r})$ is given by
\begin{equation}\label{kdens_DFTeq}
{\mathcal{H}}_{kinetic}(\vec{r}) = \frac{{\hbar}^2}{2 m} \tau_0(\vec{r}) \, ,
\end{equation}
where $m$ is the nucleon mass and $ \tau_0 = \tau_p + \tau_n $ is the total kinetic density ~\cite{dobaczewski1995time}.

With the nonlocal nuclear densities constructed, it is now possible to compute the kinetic density of a given nuclear system from \textit{ab initio} theory. We act upon the nonlocal density by a Laplacian-like operator according to the following relation described in Ref.~\cite{engel1975time},
\begin{equation}\label{kdens_eq}
{\tau}_\mathcal{N}(\vec{r}) = \bigg[ \vec{\nabla} \cdot \vec{\nabla}\,' {\rho}_\mathcal{N}(\vec{r},\vec{r}\,') \bigg]_{\vec{r}=\vec{r}\,'} \\,
\end{equation}
where $\mathcal{N}$ denotes the nucleon type for protons ($p$) and neutrons ($n$). In order to derive a computable expression for this quantity, we require several relations. It is useful to begin by writing the kinetic density in spherical component form as
\begin{equation}
\begin{split}
&\nabla_{u} \nabla_{-u}' \rho(\vec{r},\vec{r}\,') = \\
&\qquad \sum_{n,l,n',l',K,k,m_{l},m_{l'}} \alpha_{n,l,n',l'}^{K,i,f}\, (l\,m_{l}\,l'\,m_{l'} \vert LM) \\
&\qquad \times \bigg[ \nabla_{u} R_{n,l}(r) Y_{l,m_{l}}^*(\hat{r}) \bigg] \bigg[ \nabla_{-u}' R_{n',l'}(r') Y_{l',m_{l'}}^*(\hat{r}') \bigg] \quad ,
\end{split}
\end{equation}
where $ u = 0, \pm 1 $ and $ \alpha_{n,l,n',l'}^{K,i,f} $ is defined for the translationally invariant density as
\begin{equation}\label{alphaeq}
\begin{split}
&\alpha_{n,l,n',l'}^{K,i,f} = \sum_{n_1,l_1,j_1,n_2,l_2,j_2} \bigg({\frac{A}{A-1}}\,\bigg)^{3/2} \,  \\
&\qquad \times \frac{1}{\hat{J}_f} ( J_i\,M_i\,K\,k\vert J_f\,M_f) \big(M^K\big)_{n,l,n',l',n_1,l_1,n_2,l_2}^{-1} \\
&\qquad \times (-1)^{l_1+l_2+K+j_2-\frac{1}{2}} \, \hat{j}_1 \hat{j}_2 \left \{
  \begin{tabular}{ccc}
  $j_1$ & $j_2$ & $K$ \\
  $l_2$ & $l_1$ & $\frac{1}{2}$ \\
  \end{tabular}
\right \} \\
&\qquad \times {}_{SD} \langle A \lambda_f J_f \vert \vert \,(a_{n_1,l_1,j_1}^{\dagger}\,\tilde{a}_{n_2,l_2,j_2})^{(K)}\,\vert \vert A \lambda_i J_i \rangle_{SD} \, . \\
\end{split}
\end{equation}
We note that $\alpha_{n,l,n',l'}^{K,i,f}$ is different for the COM contaminated density. We now discuss several relations necessary for the derivation of the kinetic density, explicitly shown in the appendix. The first set of relations are analytic expressions for the spherical components of $ \vec{\nabla} f(\vec{r}) Y^{m}_l(\hat{r}) $, which can be found in section 5.8.3 of Ref.  \cite{varshalovich1988quantum}. In these relations, we see explicit dependence on the derivative of the RHO function. In order to remove a direct dependence on a first order differential, we introduce the following relation for the derivative of the RHO,
\begin{equation}\label{ddrR}
\frac{dR_{nl}}{dr} = \frac{l}{r} R_{nl} - \frac{1}{b} \bigg( \sqrt{n+l+\frac{3}{2}} \, R_{n,l+1}(r) +\sqrt{n} \, R_{n-1,l+1}(r) \bigg) \,.
\end{equation}
For the derivation of the Eq.~(\ref{ddrR}), see the appendix, Sec.~\ref{sec_ddrR}. Using these relations, along with additional angular momentum algebra, we may now evaluate the expression for the kinetic density in terms of spherical components, which takes the form
\begin{equation}
\begin{split}
&{\tau}_\mathcal{N}(\vec{r}) = \bigg[ \nabla_{0} \nabla_{0}' {\rho}_\mathcal{N}(\vec{r},\vec{r}\,') - \nabla_{+1} \nabla_{-1}' {\rho}_\mathcal{N}(\vec{r},\vec{r}\,') \\
&\qquad \qquad - \nabla_{-1} \nabla_{+1}' {\rho}_\mathcal{N}(\vec{r},\vec{r}\,') \bigg]_{\vec{r}=\vec{r}\,'} \, . \\
\end{split}
\end{equation}
The derivation of the $ {\nabla}_0 {\nabla}_0' {\rho}_\mathcal{N} $ component is shown in Sec.~\ref{sec_kdens}. The outlined procedure can be followed exactly for the $ {\nabla}_{+1} {\nabla}_{-1}' {\rho}_\mathcal{N} $ and $ {\nabla}_{-1} {\nabla}_{+1}' {\rho}_\mathcal{N} $ components, with only minor differences in the angular momentum algebra.

%%%% results section
\section{Results}\label{sec_results}

%%%% nonlocal density results
\subsection{Nonlocal density}\label{subsec_nlocdensres}

In this section, as in Ref.~\cite{gennari2018microscopic}, we discuss results for the nonlocal densities obtained from the NCSM wave functions using the approach described in Sec.~\ref{sec_nonlocaldensity}.

To highlight the significance of COM removal in lighter systems, we considered the $^{4,8}$He and $^{16}$O systems. We computed the translationally invariant and COM contaminated nonlocal densities, given by Eq.~(\ref{eqn:trinvnlocdens}) and Eq.~(\ref{eqn:wiCOMnlocdens}), respectively. Note that all figure plots of the COM contaminated density are labeled \textit{wiCOM} while the translationally invariant density plots are labeled \textit{trinv}. The ground state densities of the nuclei are shown with all angular dependence factorized out for plotting. Proton densities are shown in blue, neutron densities are shown in red, and total nuclear densities are shown in black.

\begin{figure}[t]
\includegraphics[width=0.5\textwidth]{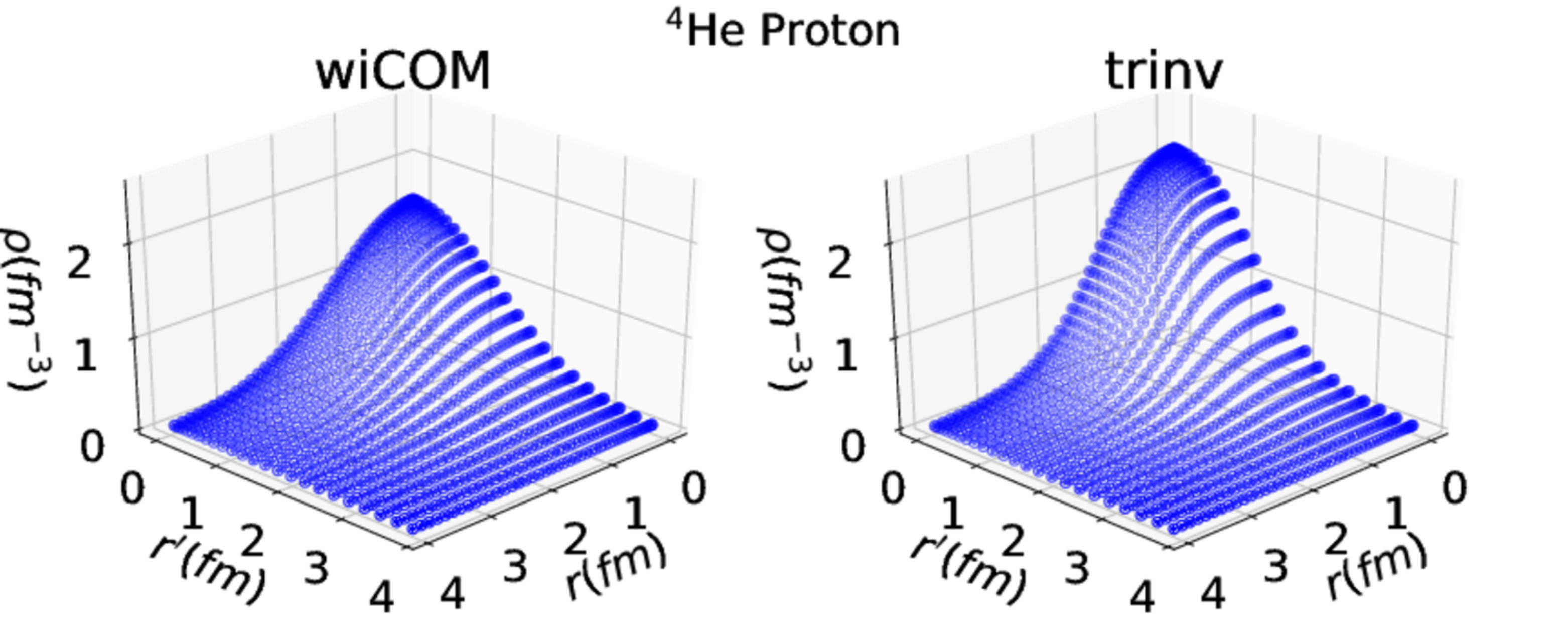}
\includegraphics[width=0.5\textwidth]{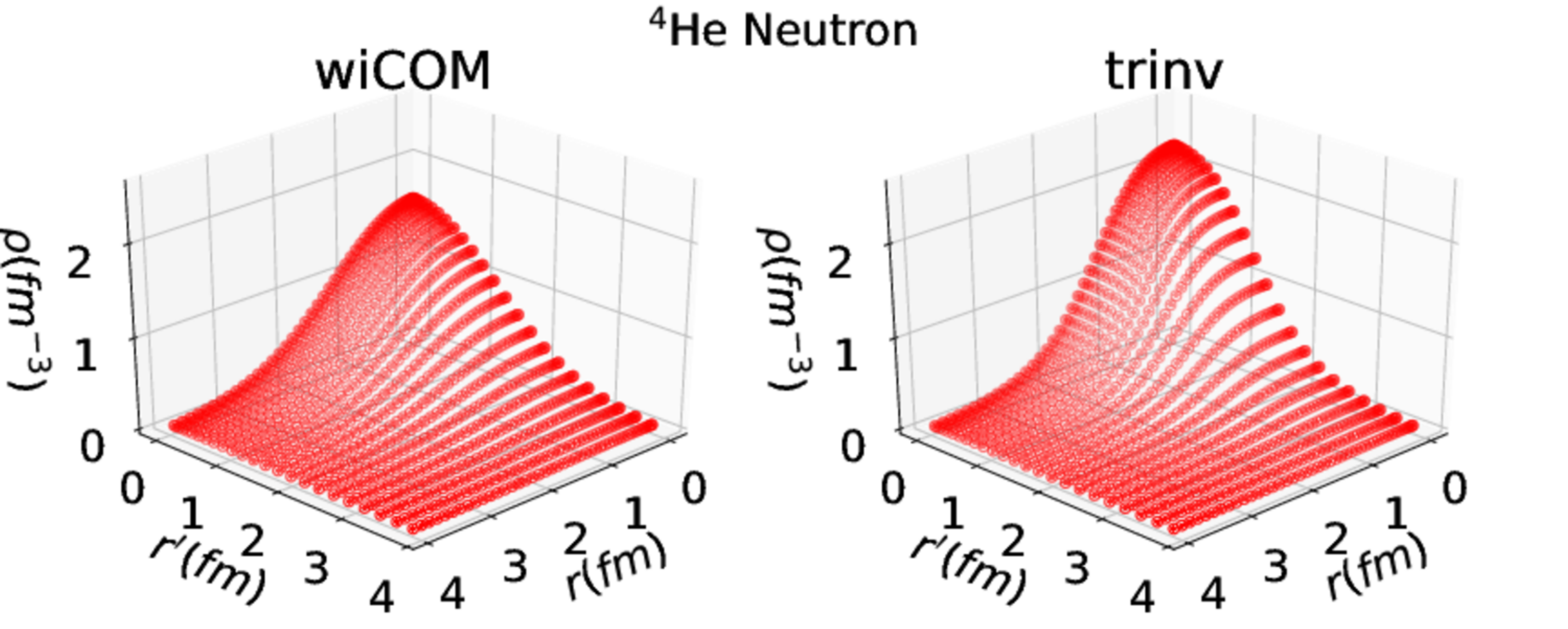}
\caption{\label{fig:nlocdens_He4bare} Ground state \textsuperscript{4}He nonlocal proton and neutron densities calculated using the bare NN-N\textsuperscript{4}LO(500) interaction with an $\nmax = 18$ basis space. An oscillator frequency of $\hb = 20.0$ MeV was used for this calculation.}
\end{figure}

\begin{figure}[t]
\includegraphics[width=0.5\textwidth]{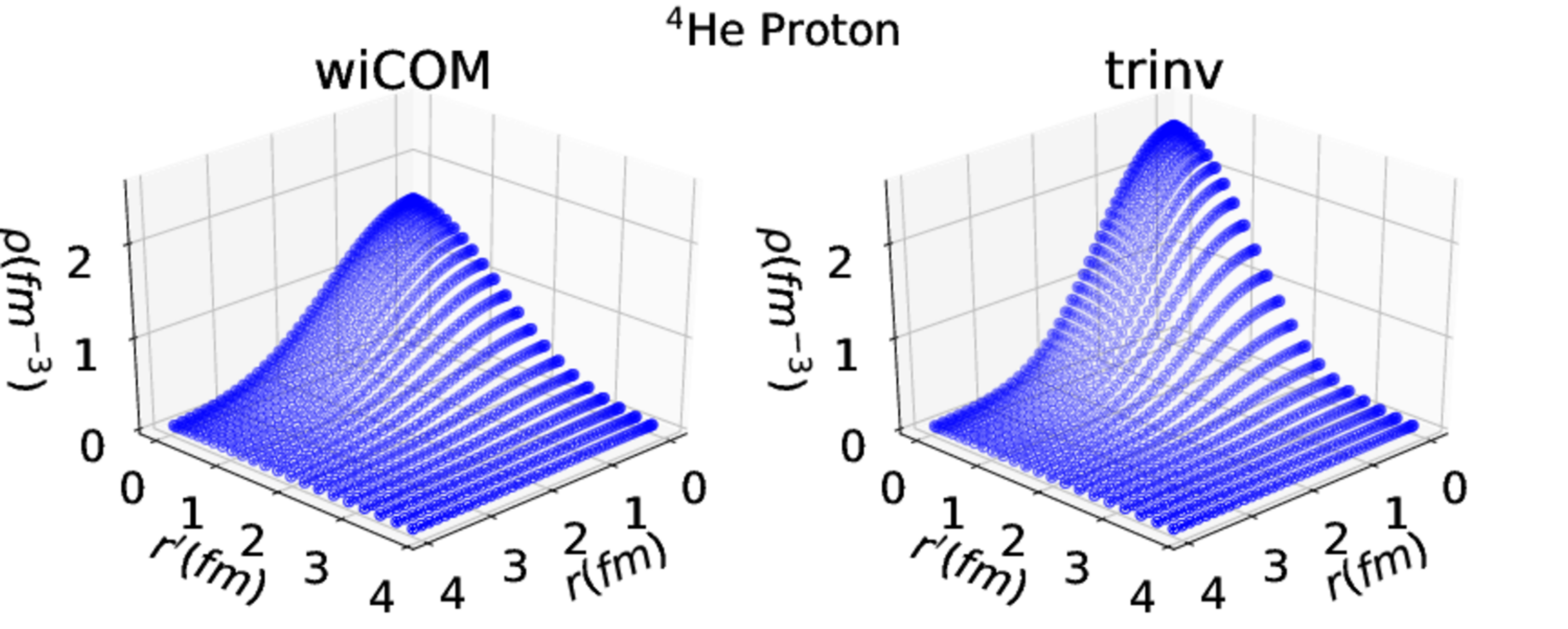}
\includegraphics[width=0.5\textwidth]{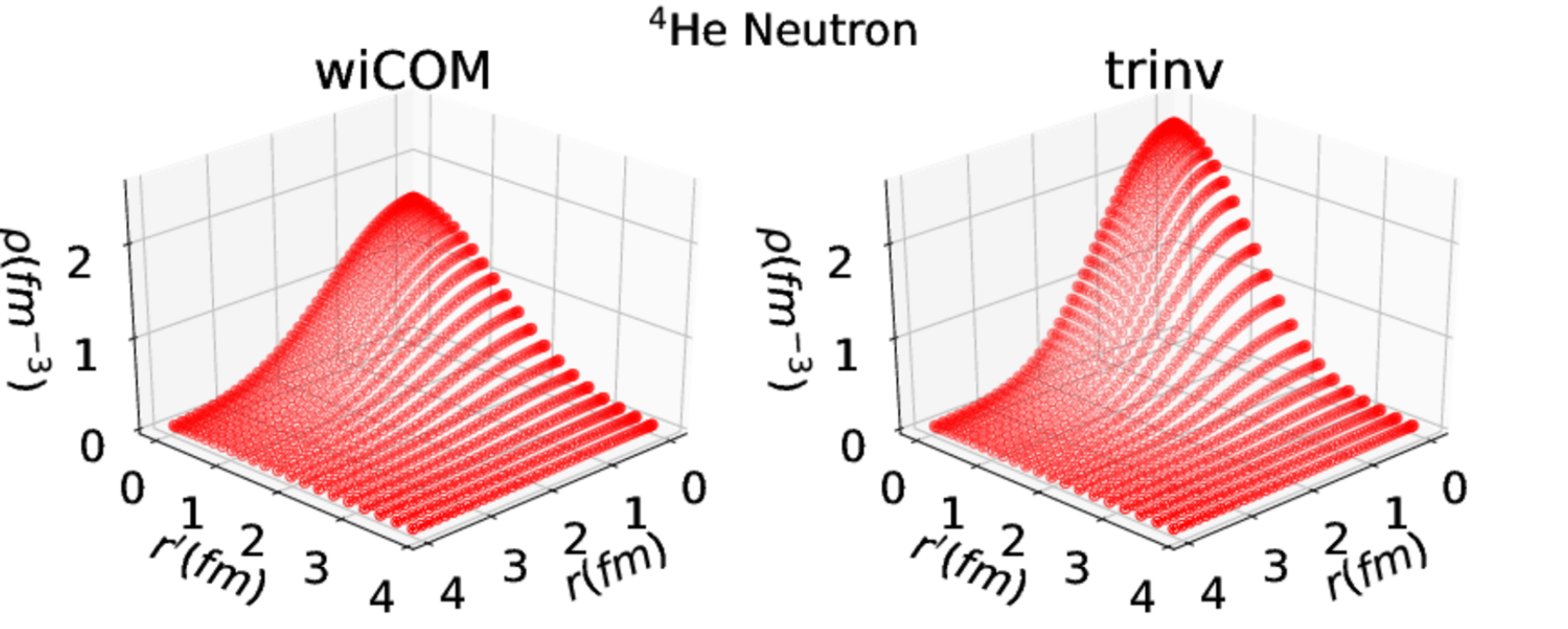}
\caption{\label{fig:nlocdens_He4} Ground state \textsuperscript{4}He nonlocal proton and neutron densities calculated using the SRG-evolved NN-N\textsuperscript{4}LO(500)+3Nlnl interaction with an $\nmax = 14$ basis space. An oscillator frequency of $\hb = 20.0$ MeV and a flow parameter of $\lambda_{\mathrm{SRG}}=2.0$ fm\textsuperscript{-1} were used for the calculations.}
\end{figure}

\begin{figure}[t]
\includegraphics[width=0.5\textwidth]{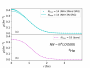}
\caption{\label{fig:bare_He4} Ground state \textit{trinv} local density comparison for \textsuperscript{4}He. \textbf{Panel a:} We show calculations with two-body (NN+3Nind SRG) and two- plus three-body (NN+3N SRG) SRG-evolved interactions. \textbf{Panel b:} We show calculations with the bare two-body NN-N\textsuperscript{4}LO(500) interaction.}
\end{figure}

In Fig.~\ref{fig:nlocdens_He4bare}, Fig.~\ref{fig:nlocdens_He4}, and Fig.~\ref{fig:bare_He4} we show comparisons between calculations of nonlocal and local $^4$He densities with the bare NN-N\textsuperscript{4}LO(500) interaction and with the previously described SRG-evolved NN-N\textsuperscript{4}LO(500)+3Nlnl interaction. The SRG-evolved interaction is computed at the two- plus three-body level, with all higher body SRG induced terms neglected. An $\nmax = 18$ basis space was used for the bare interaction, and an $\nmax = 14$ basis space was used for the SRG-evolved interaction. Comparing the nonlocal densities between the bare and SRG-evolved interactions, differences in the predicted structure are evident, such as the reduction of the peak in the bare calculation. However, an arguably more important feature of the bare density is that it tends to have an initial plateau extending to one fermi, where it begins a rapid fall off towards zero density. The rate of the declination is different from the SRG-evolved interaction, which produces a density with a smoother transition between the peak value and the fall of towards zero. These alterations are more noticeable in Fig.~\ref{fig:bare_He4}. In the top panel, we compare the local densities of the SRG-evolved two-body interaction with and without the chiral three-nucleon interaction. NN+3Nind SRG labels the two-body SRG-evolved interaction with induced three-body terms, and NN+3N SRG labels the two-body SRG-evolved interaction with the induced and chiral three-body interaction terms (full three-body interaction). Notice the differences in the density when utilizing the SRG-evolved interaction with and without the chiral three-body interaction. For all intents and purposes, we will now treat the SRG-evolved interaction as a separate, physically realistic interaction independent of the bare interaction.


\begin{figure}[t]
\includegraphics[width=0.5\textwidth]{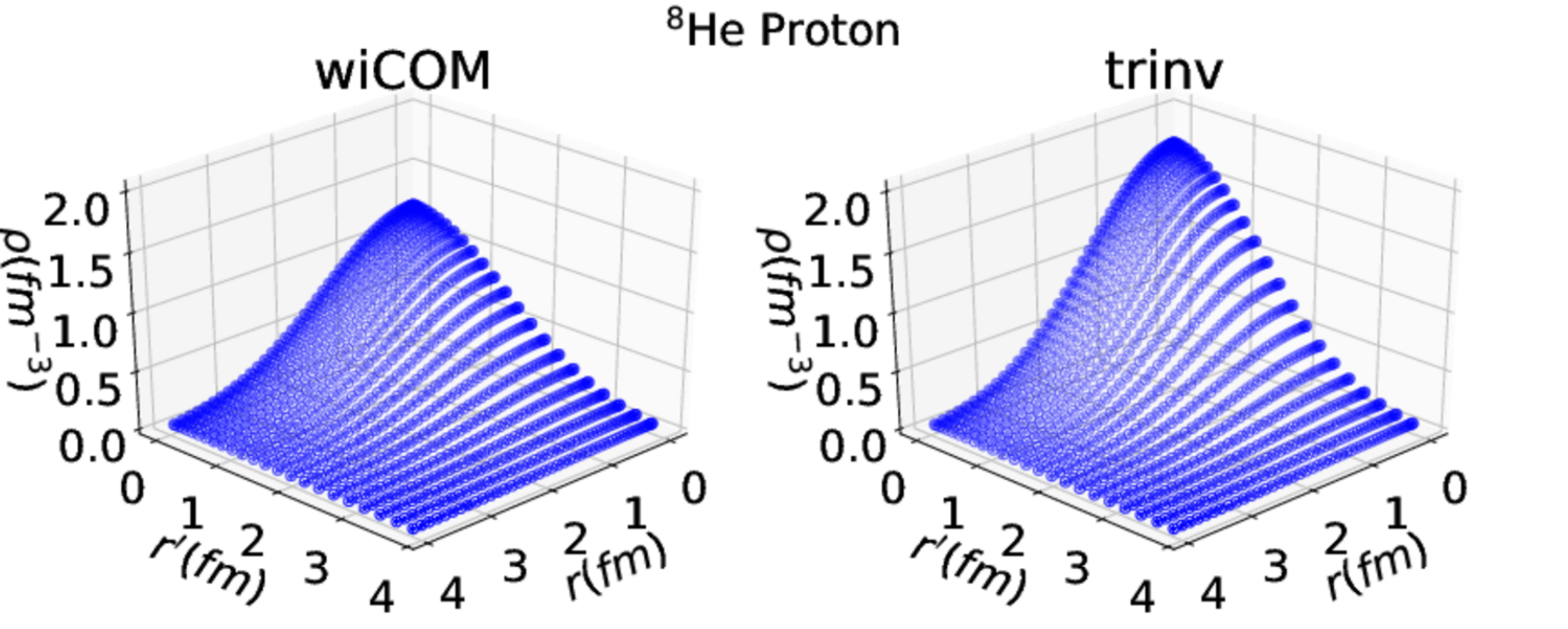}
\includegraphics[width=0.5\textwidth]{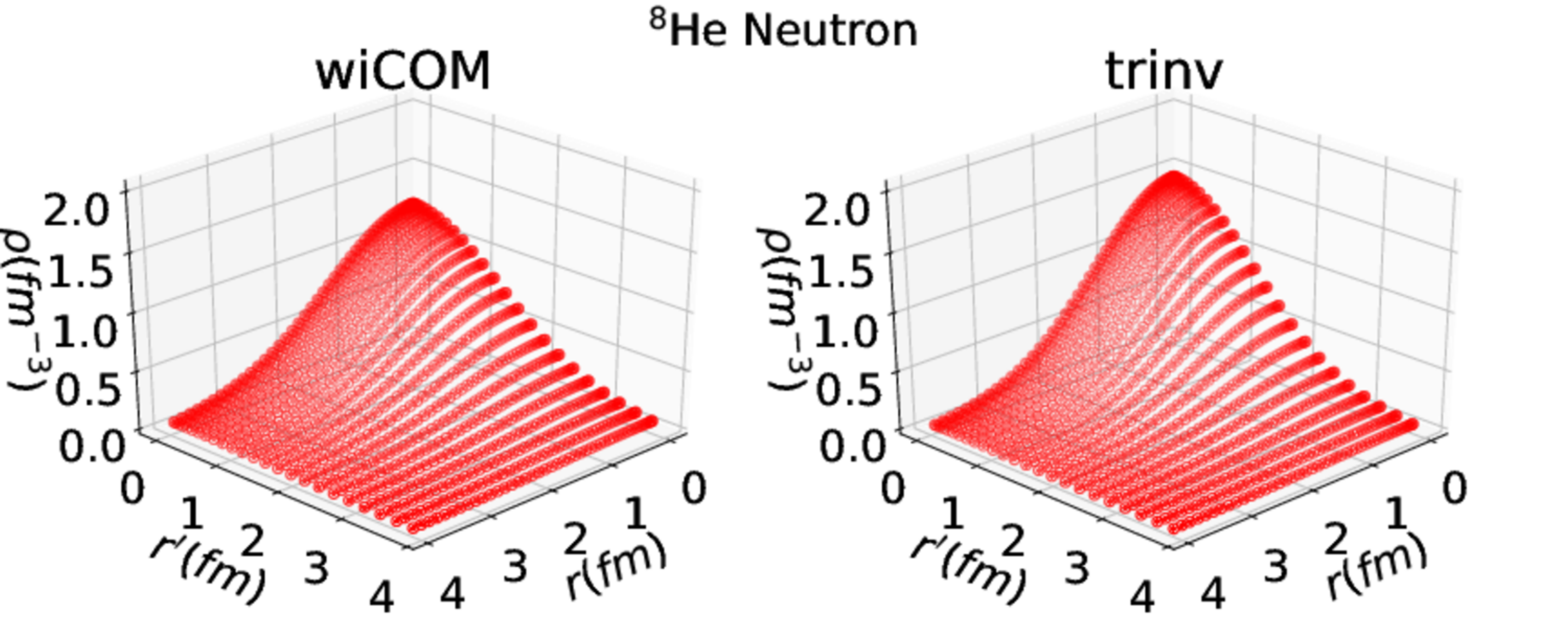}
\caption{\label{fig:nlocdens_He8} Ground state \textsuperscript{8}He nonlocal proton and neutron densities calculated using the NN-N\textsuperscript{4}LO(500)+3Nlnl interaction with an $\nmax = 10$ basis space. An oscillator frequency of $\hb = 20.0$ MeV and a flow parameter of $\lambda_{\mathrm{SRG}}=2.0$ fm\textsuperscript{-1} were used for the calculations.}
\end{figure}

In Fig.~\ref{fig:nlocdens_He4} and Fig.~\ref{fig:nlocdens_He8}, we show results for the COM contaminated and translationally invariant nonlocal proton and neutron density of $^{4,8}$He using the NN-N\textsuperscript{4}LO(500)+3Nlnl interaction. $\nmax = 14$ and $\nmax = 10$ basis spaces were used, respectively, with a flow parameter $\lambda_{\mathrm{SRG}} = 2.0$ fm\textsuperscript{-1} and an oscillator frequency of $\hb = 20.0$ MeV. To appreciate the magnitude of spurious COM contamination in light nuclei, notice the significant differences in the predicted structure of the $^{4,8}$He systems between the \textit{wiCOM} and \textit{trinv} densities. The structure differences are particularly noticeable at small $r$ and $r^{\prime}$. The \textit{trinv} has noticeably sharper features and the edges of the density tend to fall off more rapidly than in the case of the \text{wiCOM} density. Clearly there is substantial suppression of the density at small distances and a smoothing of the density over large distances coming from the COM contamination. It is also important to note the differences in density for protons and neutrons in a system such as $^{8}$He, where we see a visible smoothing of the neutron density over greater distances due to the exotic structure of the nucleus.

\begin{figure}[t]
\includegraphics[width=0.5\textwidth]{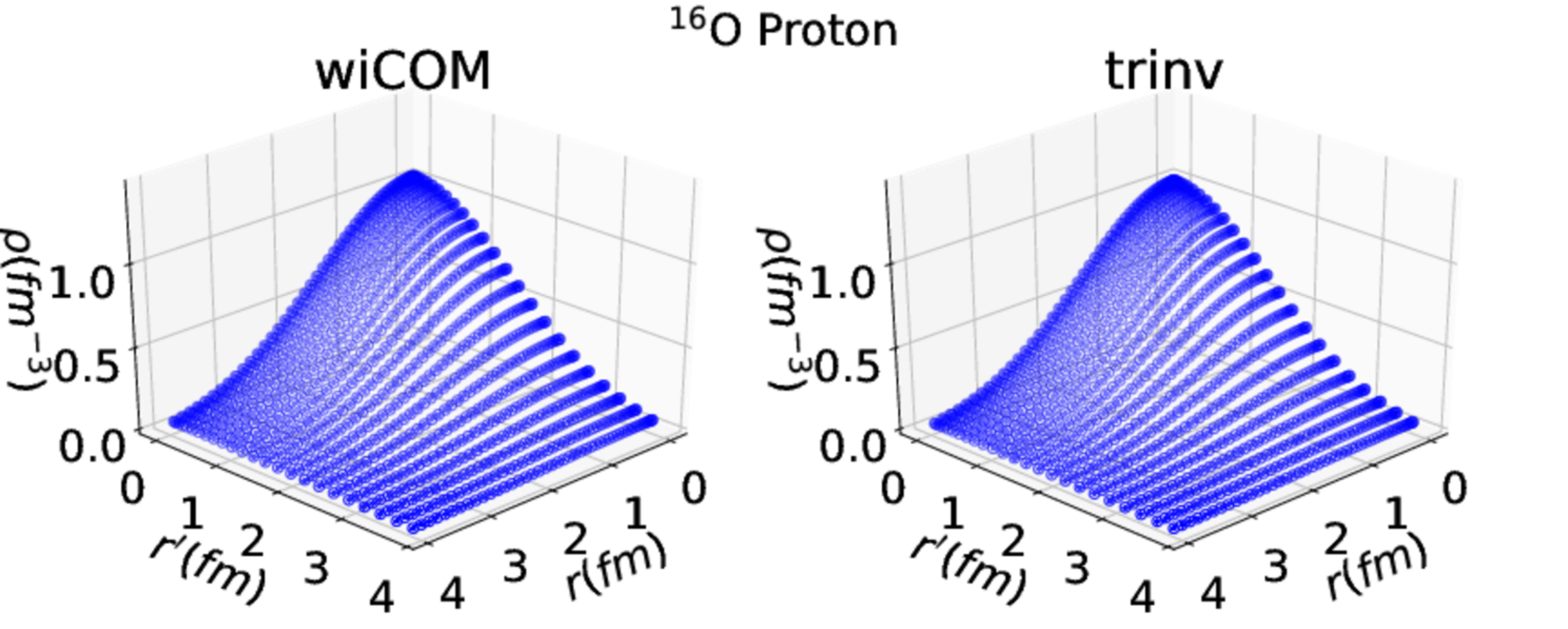}
\includegraphics[width=0.5\textwidth]{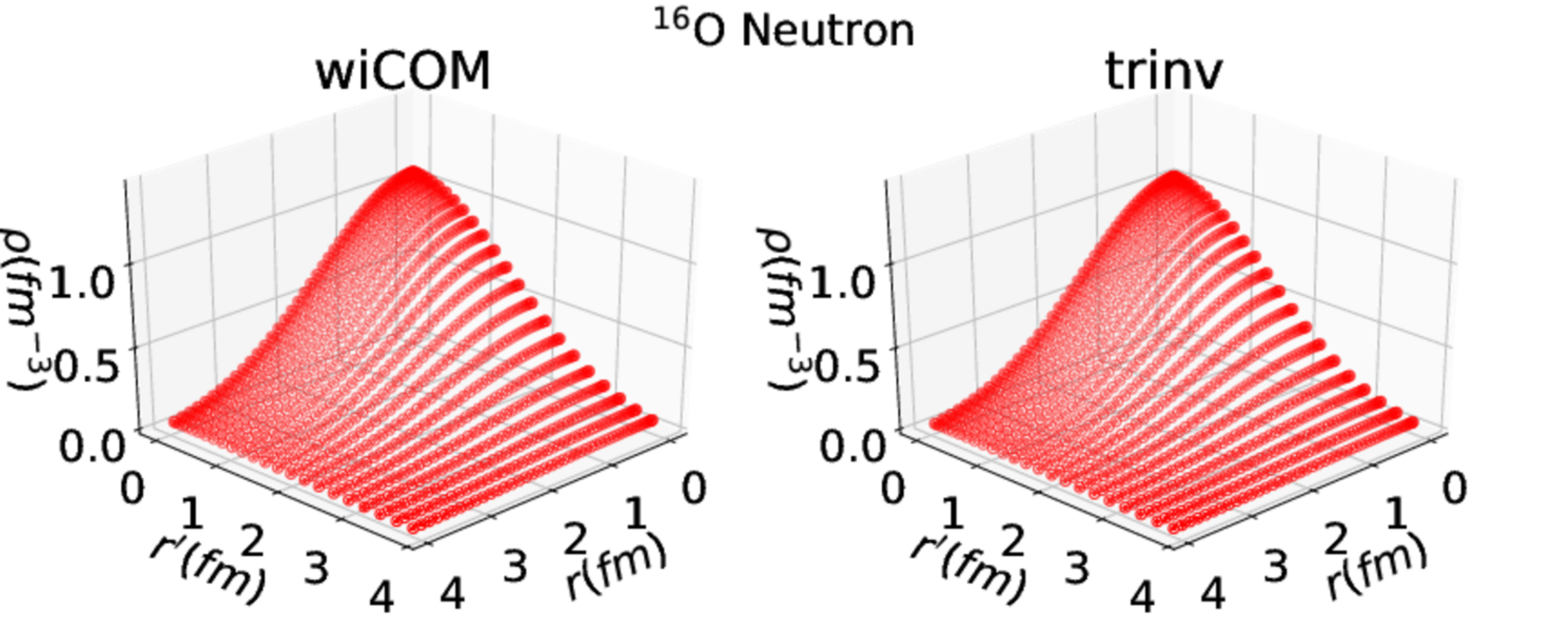}
\caption{\label{fig:nlocdens_O16} Ground state \textsuperscript{16}O nonlocal proton and neutron densities calculated using the NN-N\textsuperscript{4}LO(500)+3Nlnl interaction with an $\nmax = 8$ importance truncated basis space. An oscillator frequency of $\hb = 20.0$ MeV and a flow parameter of $\lambda_{\mathrm{SRG}}=2.0$ fm\textsuperscript{-1} were used for the calculations.}
\end{figure}

For comparison, in Fig.~\ref{fig:nlocdens_O16}, we present a similar calculation for the proton and neutron densities of $^{16}$O. There are noticeably smaller effects from the COM removal in comparison to the very light systems, $^{4,8}$He. A notable feature of the COM contamination is that it diminishes with increasing $A$, and so reduced effects are expected in larger systems. While the peaks of the \textit{trinv} density are not quite as pronounced, the smoothing effect appears to present in this larger system as the edges still fall to zero slightly more rapidly than the \textit{wiCOM} density. Nevertheless, while the COM removal effects are reduced, objects or observables highly sensitive to the structure of the density will still be impacted by these differences, as shown in Ref.~\cite{gennari2018microscopic}. One would then expect that an object such as the kinetic density, a term dependent upon a gradient on each coordinate, Eq.~(\ref{kdens_eq}), would experience an amplification of these structure differences.

\begin{figure}[t]
\includegraphics[width=0.5\textwidth]{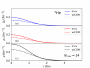}
\caption{\label{fig:locdens_He4} Ground state \textsuperscript{4}He local proton (\textbf{panel a}), neutron (\textbf{panel b}), and total densities (\textbf{panel c}) computed using the NN-N\textsuperscript{4}LO(500)+3Nlnl interaction with an $\nmax = 14$ basis space, an oscillator frequency of $\hb = 20.0$ MeV, and a flow parameter of $\lambda_{\mathrm{SRG}}=2.0$ fm\textsuperscript{-1}.}
\end{figure}
\begin{figure}[t]
\includegraphics[width=0.5\textwidth]{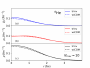}
\caption{\label{fig:locdens_He8} Ground state \textsuperscript{8}He local proton (\textbf{panel a}), neutron (\textbf{panel b}), and total densities (\textbf{panel c}) computed using the NN-N\textsuperscript{4}LO(500)+3Nlnl interaction with an $\nmax = 10$ basis space, an oscillator frequency of $\hb = 20.0$ MeV, and a flow parameter of $\lambda_{\mathrm{SRG}}=2.0$ fm\textsuperscript{-1}.}
\end{figure}

\begin{figure}[t]
\includegraphics[width=0.5\textwidth]{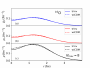}
\caption{\label{fig:locdens_O16} Ground state \textsuperscript{16}O local proton (\textbf{panel a}), neutron (\textbf{panel b}), and total densities (\textbf{panel c}) computed using the NN-N\textsuperscript{4}LO(500)+3Nlnl interaction with an $\nmax = 8$ importance truncated basis space, an oscillator frequency of $\hb = 20.0$ MeV, and a flow parameter of $\lambda_{\mathrm{SRG}}=2.0$ fm\textsuperscript{-1}.}
\end{figure}

We now present the local proton and neutron densities, $\rho_{\mathcal{N}}(r) = \rho_{\mathcal{N}}(r,r)$, for $^{4,8}$He and $^{16}$O for further analysis. Referring to Fig.~\ref{fig:locdens_He4} and Fig.~\ref{fig:locdens_He8} for the local densities of light nuclei, there are notably drastic effects resulting from the COM removal procedure. If accurate nuclear structure calculations are to be performed for lighter systems, one must properly treat the COM contamination in these systems. Additionally, in studying the local densities of $^{16}$O in Fig.~\ref{fig:locdens_O16}, one can see structural differences present in the larger system which were not so easily observed in the nonlocal density figures. From the local densities we observe that these structure differences are apparent and still relevant in the larger systems, even though the COM contribution diminishes with increasing $A$-nucleon number. As a result, we expect that the COM removal process will produce noticeable changes in the kinetic densities for $^{16}$O.

%

%%%% kinetic density results
\subsection{Kinetic density}\label{subsec_kdensres}
%re-add He6
In the following section we present the main result of this work; kinetic densities computed from \textit{ab initio} NCSM nonlocal densities using the method outlined in Sec.~\ref{sec_kineticdensity}. For completeness, we present results ranging from $^{4}$He to $^{16}$O, though we emphasize that any reasonable comparison with DFT can only be done with the latter.

\begin{figure}[t]
\includegraphics[width=0.5\textwidth]{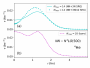}
\caption{\label{fig:bare_He4kdens} Ground state \textit{trinv} kinetic density comparison for \textsuperscript{4}He. \textbf{Panel a:} We show calculations with two-body (NN+3Nind SRG) and two- plus three-body (NN+3N SRG) SRG-evolved interactions. \textbf{Panel b:} We show calculations with the bare two-body NN-N\textsuperscript{4}LO(500) interaction. Nonlocal densities were computed as previously described in Fig.~\ref{fig:nlocdens_He4bare} and Fig.~\ref{fig:nlocdens_He4}, respectively.}
\end{figure}

As for the densities, we present results for $^{4}$He using SRG-evolved chiral two-body (NN+3Nind SRG) and chiral two- plus three-body (NN+3N SRG) interactions in the top panel, as well as the bare NN-N\textsuperscript{4}LO(500) interaction kinetic densities in the bottom panel of Fig.~\ref{fig:bare_He4kdens}. As previously discussed, we see significant differences with the inclusion of the chiral three-body interaction terms when using the SRG-evolved interaction to compute the kinetic density. The most significant differences in the predicted structure of the SRG-evolved and bare interactions occur at ranges of less than one fermi. To reiterate, moving forward we will treat the SRG-evolved NN-N\textsuperscript{4}LO(500)+3Nlnl interaction as a different, physically realistic interaction.

\begin{figure}[t]
\includegraphics[width=0.5\textwidth]{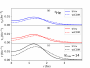}
\caption{\label{fig:kdens_He4} Ground state \textsuperscript{4}He comparisons between the \textit{trinv} and \textit{wiCOM} kinetic densities. Proton (\textbf{panel a}), neutron (\textbf{panel b}), and total kinetic densities (\textbf{panel c}) are shown. The nonlocal density was computed as previously described in Sec.~\ref{subsec_nlocdensres}. The expectation value of the intrinsic kinetic energy for $^{4}$He is $51.91$ MeV.}
\end{figure}

\begin{figure}[t]
\includegraphics[width=0.5\textwidth]{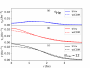}
\caption{\label{fig:kdens_He6} Ground state \textsuperscript{6}He comparisons between the \textit{trinv} and \textit{wiCOM} kinetic densities. Proton (\textbf{panel a}), neutron (\textbf{panel b}), and total kinetic densities (\textbf{panel c}) are shown. The nonlocal density was computed as previously described in Sec.~\ref{subsec_nlocdensres}. The expectation value of the intrinsic kinetic energy for $^{6}$He is $78.26$ MeV.}
\end{figure}

\begin{figure}[t]
\includegraphics[width=0.5\textwidth]{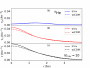}
\caption{\label{fig:kdens_He8} Ground state \textsuperscript{8}He comparisons between the \textit{trinv} and \textit{wiCOM} kinetic densities. Proton (\textbf{panel a}), neutron (\textbf{panel b}), and total kinetic densities (\textbf{panel c}) are shown. The nonlocal density was computed as previously described in Sec.~\ref{subsec_nlocdensres}. The expectation value of the intrinsic kinetic energy for $^{8}$He is $116.30$ MeV.}
\end{figure}

Let us now consider the lighter systems to gauge the significance of the COM removal process, and to understand how the COM contamination may impact objects dependent on the nonlocal density. In Figs.~\ref{fig:kdens_He4}, ~\ref{fig:kdens_He6} and ~\ref{fig:kdens_He8}, we present results for the kinetic density of $^{4,6,8}$He, respectively, with the nonlocal proton and neutron densities computed as previously described in Sec.~\ref{subsec_nlocdensres}. As expected, for small $A$-nucleon systems we observe tremendous differences in the \textit{trinv} and \textit{wiCOM} kinetic densities, most significantly in the case of $^{4}$He. The amplification of the density structure differences is quite pronounced, and further we see the suppression previously attributed to the COM contamination appearing in the kinetic densities. We find the maximum suppression occurring in the short range distances, while the COM contamination tends to spread the kinetic densities over larger distances, as was observed for the nonlocal densities. Notice that not only do we see significant differences with small $r$, but we see fairly pronounced changes in the long range behavior of the kinetic density in nuclei like $^{6,8}$He.


\begin{figure}[t]
\includegraphics[width=0.5\textwidth]{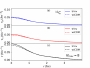}
\caption{\label{fig:kdens_C12} Ground state \textsuperscript{12}C comparisons between the \textit{trinv} and \textit{wiCOM} kinetic densities. Proton (\textbf{panel a}), neutron (\textbf{panel b}), and kinetic total densities (\textbf{panel c}) are shown. The nonlocal density was computed as previously described in Sec.~\ref{subsec_nlocdensres}. The expectation value of the intrinsic kinetic energy for $^{12}$C is $219.84$ MeV.}
\end{figure}

\begin{figure}[t]
\includegraphics[width=0.5\textwidth]{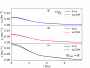}
\caption{\label{fig:kdens_O16} Ground state \textsuperscript{16}O comparisons between the \textit{trinv} and \textit{wiCOM} kinetic densities. Proton (\textbf{panel a}), neutron (\textbf{panel b}), and total kinetic densities (\textbf{panel c}) are shown. The nonlocal density was computed as previously described in Sec.~\ref{subsec_nlocdensres}. The expectation value of the intrinsic kinetic energy for $^{16}$O is $301.69$ MeV.}
\end{figure}

Let us now consider heavier $A$-nucleon systems, which can provide a method of directly gauging the impact of COM contamination in energy density functionals. In Fig.~\ref{fig:kdens_C12} and Fig.~\ref{fig:kdens_O16}, we present the results for the kinetic density of $^{12}$C, with the nonlocal densities computed as previously described in Sec.~\ref{subsec_nlocdensres}. As expected from previous results, the \textit{wiCOM} nonlocal densities suppress the kinetic density for small $r$, however the effect is not as pronounced as in the lighter systems of $^{4,6,8}$He.

In both the $^{12}$C and $^{16}$O systems, we see significantly reduced effects during the COM removal process. This may be in part due to the highly spherical shape of a system such as $^{16}$O, though this requires further inspection.  While reduced, the \textit{trinv} and \textit{wiCOM} nonlocal densities maintain a non-negligible difference which may provide some corrections if used as an input for energy density functionals.

\begin{figure}[t]
\includegraphics[width=0.5\textwidth]{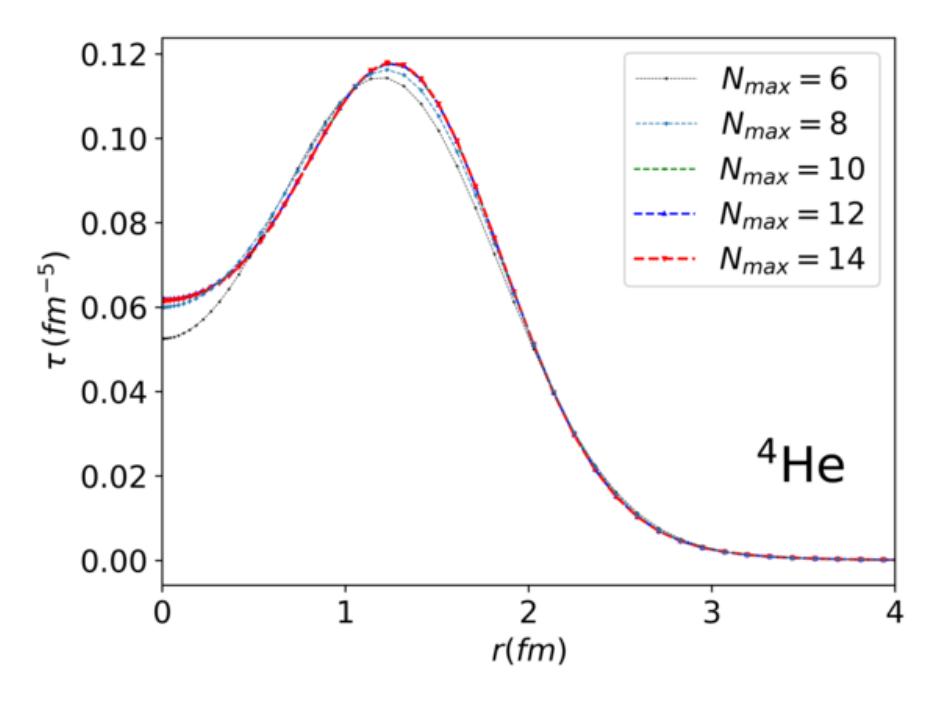}
\caption{\label{fig:kdensnmax_He4} Ground state $\nmax$ convergence results for \textsuperscript{4}He \textit{trinv} kinetic neutron density. The nonlocal density was computed as previously described in Sec.~\ref{subsec_nlocdensres}.}
\end{figure}
\begin{figure}[t]
\includegraphics[width=0.5\textwidth]{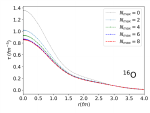}
\caption{\label{fig:kdensnmax_O16} Ground state $\nmax$ convergence results for \textsuperscript{16}O \textit{trinv} kinetic neutron density. The nonlocal density was computed as previously described in Sec.~\ref{subsec_nlocdensres}.}
\end{figure}
Let us now turn to a discussion on the integration of the kinetic density operator. Note that in this work we consider solely the $J=0$ ground states. In the case of the translationally invariant kinetic density, upon integration over the spatial coordinates, we exactly reproduce the expectation value of the ground state intrinsic kinetic energy of the nucleus, which can be independently calculated from two-body densities introduced in second quantization. The expectation value is given by Eq. (\ref{eq:tbdTint}),
\begin{equation}\label{eq:tbdTint}
\begin{split}
& \langle T_{int} \rangle = \frac{1}{4} \sum_{abcd} \bra{ a b } T_{int} \ket{ c d } \\
& \qquad \, \; \times {}_{SD}\bra{A \lambda J T} a^{\dagger}_a a^{\dagger}_b a_d a_c \ket{A \lambda J T}{}_{SD}  \, . \\
\end{split}
\end{equation}
\begin{table}
\setlength{\tabcolsep}{0.75em}
\def\arraystretch{1.75}
\begin{tabular}{ | c | c | c | c | }
 \hline
 Nucleus & $\nmax$ &\; $\langle T_{int} \rangle \; $ & Error ($\pm$) \\ [1ex]
 \hline
 $^{4}$He (bare) & 18 & 62.73 & $\pm$ 0.01 \% \\ [1ex]
 \hline
 $^{4}$He & 14 & 51.91 & $\pm$ 0.01 \% \\ [1ex]
 \hline
  $^{6}$He & 12 & 78.27 & $\pm$ 1.4 \% \\ [1ex]
 \hline
 $^{8}$He & 10 & 116.30 & $\pm$ 3.1 \% \\ [1ex]
 \hline
 $^{12}$C & 8 IT& 219.84 & $\pm$ 1.2 \% \\ [1ex]
 \hline
 $^{16}$O & 8 IT & 301.69 & $\pm$ 0.8 \% \\  [1ex]
 \hline
\end{tabular}
\caption{\label{tbl:Tint} Ground state mean intrinsic kinetic energy values and percent errors for all aforementioned nuclei calculated using the NN-N\textsuperscript{4}LO(500)+3Nlnl interaction (except $^{4}$He-bare, which are the results for the bare NN-N\textsuperscript{4}LO(500) interaction). All $\langle T_{int} \rangle $ values are in MeV. Note \textit{IT} refers to an importance truncated basis space. Percent errors are calculated using the difference between the maximal $\nmax$ value and the previous value.}
\end{table}
When considering the COM contaminated kinetic density, one recovers the expectation value of the intrinsic kinetic energy plus the expectation value of the kinetic energy of the COM. The results for the $ \langle T_{int} \rangle $ are summarized in Table~\ref{tbl:Tint}. The recovery of the intrinsic kinetic energy after COM removal is direct confirmation of success of the procedure, and can be summarized by the following set of relations, Eq.~(\ref{eq:Ttrinv}) and Eq.~(\ref{eq:TwiCOM}),
\begin{equation}\label{eq:Ttrinv}
\begin{split}
& \langle T_{int} \rangle = {}_{SD}\bra{A \lambda J T} \bigg( \frac{{\hbar}^2}{2 m } \tau^{\textit{trinv}}_{0} \bigg) \ket{A \lambda J T}{}_{SD} \\
& \qquad \, \; = \frac{{\hbar}^2}{2 m } \int_0^{\infty} r^2 \tau^{trinv}_{0}(r) \, dr \; , \\
\end{split}
\end{equation}
\begin{equation}\label{eq:TwiCOM}
\begin{split}
& \langle T_{wiCOM} \rangle = {}_{SD}\bra{A \lambda J T} \bigg( \frac{{\hbar}^2}{2 m } \tau^{\textit{wiCOM}}_{0} \bigg) \ket{A \lambda J T}{}_{SD} \\
& \qquad \qquad \, \, = \frac{{\hbar}^2}{2 m } \int_0^{\infty} r^2 {\tau}_{0}^{wiCOM}(r) \, dr \\
& \qquad \qquad \, \, = {}_{SD}\bra{A \lambda J T} \frac{{\hbar}^2}{2 m } \bigg ( \tau^{\textit{int}}_{0} + \tau^{\textit{COM}}_{0} \bigg ) \ket{A \lambda J T}{}_{SD} \\
& \qquad \qquad \, \, = \langle T_{int} \rangle + \frac{3}{4} \hb \; , \\
\end{split}
\end{equation}
where $m$ is the nucleon mass and ${\tau}_0$ is the total kinetic density. Note that these relations are always true in the NCSM, whereas in other methods are only true if convergence to an exact many-body solution is achieved. In Fig.~\ref{fig:kdensnmax_He4} and Fig.~\ref{fig:kdensnmax_O16}, we present ground state $\nmax$ convergence plots for the kinetic density of the nuclei $^{4}$He and $^{16}$O. We achieve rapid convergence in $^{4}$He when applying the NN-N4LO(500)+3Nlnl interaction at a basis size of $\nmax = 10$, as the final three $\nmax$ calculations ($\nmax=10,12,14$) overlap completely. Similarly, we are able to see good convergence trends in $^{16}$O at an importance truncated basis size of $\nmax = 8$, as this calculation is only mildly different from the the $\nmax = 6$ basis space calculation. Let it be noted that given our use of the harmonic oscillator basis, all densities - and density dependent quantities - have ``unphysical'' asymptotic behaviour due to the Gaussian tail resulting from the basis expansion.
\begin{center}
\begin{figure*}[t]
\includegraphics[width=0.75\textwidth]{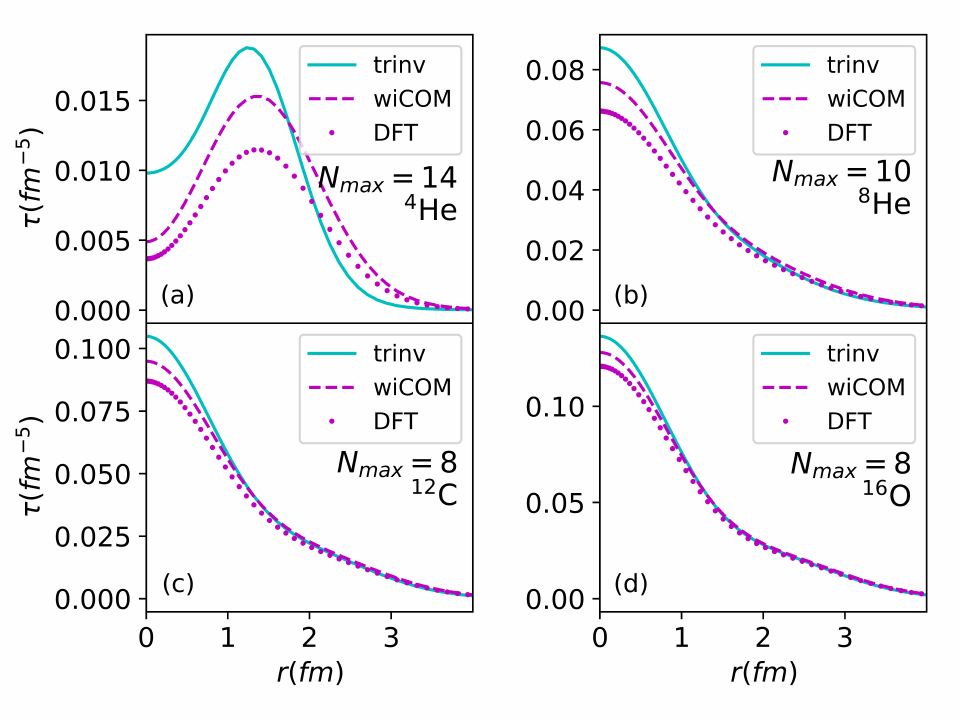}
\caption{\label{fig:kdensDFTcomp} Ground state total kinetic density results for \textsuperscript{4}He (\textbf{panel a}), \textsuperscript{8}He (\textbf{panel b}), \textsuperscript{12}C (\textbf{panel c}), and \textsuperscript{16}O (\textbf{panel d}) calculated with the NN-N\textsuperscript{4}LO(500)+3Nlnl interaction. The nonlocal densities for the nuclei were computed as previously described in Sec.~\ref{subsec_nlocdensres}. The DFT kinetic density was obtained by using Eq.~(\ref{kdens_eq_rmCOM}), $ \tau_{DFT}(r) = (1-\frac{1}{A}) \tau_{wiCOM}(r) $.}
\end{figure*}
\end{center}
%

%%%% comparisons to DFT predictions
\subsection{Comparison to basic COM treatment in DFT}\label{subsec_compDFT}

Let us now revisit the form of Eq.~(\ref{kdens_DFTeq}). This $\mathcal{H}_{kinetic}$ term has no additional treatment for the COM contamination. However, a basic COM treatment can be introduced in DFT \cite{dobaczewski2003nuclear,bender2003self,bender2000consequences}. In Eq.~(\ref{kdens_DFTeq_rmCOM}), a term inversely proportional to the number of nucleons is subtracted from the standard $\mathcal{H}_{kinetic}$ to treat the COM contamination:
\begin{equation}\label{kdens_DFTeq_rmCOM}
{\mathcal{H}}_{kinetic}(\vec{r}) = \frac{{\hbar}^2}{2 m} \bigg( 1 - \frac{1}{A} \bigg) \tau_0(\vec{r}) \; ,
\end{equation}
where $\tau_0$ would be $\tau_{wiCOM}$ in our calculations. In Fig.~\ref{fig:kdensDFTcomp}, we show \textit{trinv}, \textit{wiCOM}, and DFT calculations of the kinetic density for $^{4,8}$He, $^{12}$C, and $^{16}$O, obtained using the NN-N\textsuperscript{4}LO(500)+3Nlnl interaction. The DFT curve is obtained by application of  Eq.~(\ref{kdens_DFTeq_rmCOM}), so
\begin{equation}\label{kdens_eq_rmCOM}
\tau_{DFT}(\vec{r}) = \bigg(1-\frac{1}{A}\bigg) \tau_{wiCOM}(\vec{r}) \; .
\end{equation}
The most important item to note about the plots is the difference in the kinetic density profile when comparing the \textit{ab initio} calculation to the mock DFT calculation, which includes the aforementioned COM treatment. The differences between the predicted kinetic density structure are easier seen in the lighter nuclei, where the effects of COM removal are more drastic. Nevertheless these effects are still appreciable in the larger systems under consideration. The DFT calculation including the COM treatment has reduced the overall size of the \textit{wiCOM} kinetic density significantly. In particular, the inclusion of this $\frac{1}{A}$ term pushes the short range segments of the DFT curve further from the \textit{ab initio} translationally invariant kinetic density, whereas the long range portions are pushed closer. As expected, with increasing nucleon number the total change from the \textit{wiCOM} kinetic density is reduced, yet still non-negligible in a system such as \textsuperscript{16}O. This comparison has shown that perhaps this first order COM correction employed in DFT is an inaccurate method of treating contamination in low-mass nuclei, pushing the structure of the kinetic density further from the \textit{ab initio} prediction. While this was noted as a potential source of error of the technique, inducing slightly unphysical trends with respect to the nucleon number~\cite{bender2003self,bender2000consequences}, there have been few efforts to quantify the magnitude of the effect with respect to \textit{ab initio} calculations. As was the purpose of this work, we may now directly compare NCSM and DFT kinetic density results and, as it is not itself an observable, we may compare the effects that these structural changes will have on the minimization of the energy density functional. We anticipate that these COM corrections in the kinetic density will produce fine structure corrections to ground state observables.

In Table~\ref{tbl:TintCOMP}, we present the mean kinetic energy values for the \textit{trinv}, \textit{wiCOM}, and DFT calculations. Comparing the $ \langle T_{int} \rangle $ and $ \langle T_{DFT} \rangle $ columns, one can see that the mean values agree well across both COM removal techniques, with the $ \langle T_{DFT} \rangle $ consistently slightly underestimating the true value of the mean. Notably, the increasing nucleus size increases the difference between the \textit{ab initio} and DFT mean intrinsic kinetic energy values, indicating some form of unphysical trend with respect to the $A$-nucleon number in the DFT calculation. The inclusion of this $\frac{1}{A}$ term in the DFT calculation does appear to reduce the integral of the kinetic density appropriately, effectively removing spurious COM contamination from the mean value intrinsic kinetic energy, albeit with a very different structural prediction for the kinetic density.
\begin{table}[t!]
\setlength{\tabcolsep}{0.75em}
\def\arraystretch{1.75}
\begin{tabular}{ | c | c | c | c | c |}
 \hline
 Nucleus & $\nmax$ & $ \langle T_{int} \rangle $ &  $ \langle T_{wiCOM} \rangle $ & $ \langle T_{DFT} \rangle $ \\ [1ex]
 \hline
 $^{4}$He & 14 & 51.91 & 66.91 & 50.18 \\ [1ex]
 \hline
 $^{6}$He & 12 & 78.26 & 93.26 & 77.72 \\ [1ex]
 \hline
 $^{8}$He & 10 & 116.30 & 131.30 & 114.89 \\ [1ex]
 \hline
 $^{12}$C & 8 IT & 219.84 & 234.84 & 215.27 \\ [1ex]
 \hline
 $^{16}$O & 8 IT & 301.69 & 316.69 & 296.90 \\  [1ex]
 \hline
\end{tabular}
\caption{\label{tbl:TintCOMP} Ground state mean intrinsic kinetic energy values using \textit{trinv}, \textit{wiCOM}, and DFT kinetic densities for all aforementioned nuclei, calculated with the NN-N\textsuperscript{4}LO(500)+3Nlnl interaction. All $\langle T_i \rangle $ values are in MeV. Note \textit{IT} refers to an importance truncated basis space. The $ \langle T_{DFT} \rangle $ is calculated by using Eq.~(\ref{kdens_eq_rmCOM}), $ \langle T_{DFT} \rangle = (1-\frac{1}{A}) \langle T_{wiCOM} \rangle $. The values of $\langle T_{int} \rangle$ and $\langle T_{wiCOM} \rangle$ differ just by $\textstyle\frac{3}{4}\hbar\Omega$, see Eq.~(\ref{eq:TwiCOM}).}
\end{table}
%

%%%% conclusions
\section{Conclusions}\label{sec_conclusions}

Motivated by the recent efforts to connect DFT and \textit{ab initio} approaches to the nuclear many-body problem, the purpose of this work was to provide \textit{ab initio} predictions for the nuclear kinetic density, a fundamental input of energy density functionals in DFT, such that comparisons can then be produced for both the many-body methods and the COM removal techniques. We used the approach of Ref.~\cite{gennari2018microscopic} to construct both COM contaminated and translationally invariant nonlocal one-body densities. The kinetic densities were then computed following the procedure outlined in Sec.~\ref{sec_kineticdensity}, which provided an analytic expression in terms of the one-body density matrix elements that was then evaluated numerically. The nuclear density and kinetic density results were obtained using the SRG-evolved NN-N\textsuperscript{4}LO(500)+3Nlnl chiral interaction~\cite{gennari2018microscopic, entem2017high, navratil2007local}.

The calculation of the one-body density matrix elements and nonlocal densities requires the knowledge of the many-body nuclear wave functions, which in this work were computed from the {\it ab initio} NCSM approach. In Sec.~\ref{subsec_nlocdensres}, we showed results with and without the ground state COM contamination for the densities of $^{4,8}$He and $^{16}$O, obtained from the NCSM wave functions. As observed in the Sec.~\ref{subsec_kdensres}, the COM removal process produces non-negligible structure changes in both the nonlocal densities and, further, in the kinetic densities of $^{4,6,8}$He, $^{12}$C and $^{16}$O. In Sec.~\ref{subsec_compDFT}, we performed a comparison of the \textit{trinv} kinetic density to a basic COM removal technique used in DFT~\cite{dobaczewski2003nuclear,bender2003self,bender2000consequences}. While the COM treatment provided good agreement for the mean value intrinsic kinetic energy of the nuclei, the DFT kinetic density was shown to be structurally different from the \textit{ab initio} calculations, forcing long-range behavior closer to and short-range behavior further from the NCSM result. Comparisons such as this provide insight into refining energy density functionals, perhaps providing fine structure corrections to ground state observables in DFT. It should be noted that given the latest DFT developments, e.g., in Ref.~\cite{microscopicEDF2018}, where it is attempted to derive the energy density functionals from chiral forces, a direct comparison to our calculations will be possible as exactly the same chiral forces can be used as input in both types of calculations.

In conclusion, the development of a general nonlocal density allows for the calculation of fundamental quantities taken as input in theories such as DFT. This provides the communities with a means to better gauge the differences in many-body techniques and procedures for COM removal. Although the COM removal effect is reduced in larger $A$-nucleon systems, it is still non-negligible and can motivate the need to include a procedural technique for removing the COM or motivate a check against the existing techniques of COM removal.

%%%% appendix
\begin{widetext}
\section{Appendix}\label{sec_appendix}

%%%% derivative of radial harmonic oscillator function (RHO)
\subsection{Derivative of radial harmonic oscillator function}\label{sec_ddrR}
To begin, we introduce existing derivative and recurrence relations for Laguerre polynomials:
\begin{equation}
\label{derLaguerre}
\frac{d}{dr} L_{n}^{l}(r) = - L_{n-1}^{l+1}(r)
\end{equation}
\begin{equation}
\label{sumLaguerre}
L_{n}^{l}(r) + L_{n-1}^{l+1}(r) = L_{n}^{l+1}(r)
\end{equation}
Recall that the radial harmonic oscillator (RHO) function is given by
\begin{equation}
R_{n,l}(r) = \sqrt{\frac{2\Gamma(n+1)}{(b^2)^{l+\frac{3}{2}}\Gamma(n+l+\frac{3}{2})}}\,r^{l} \exp\Big(-\frac{r^2}{2b^2}\Big)\,L_{n}^{l+\frac{1}{2}}\bigg(\frac{r^2}{b^2}\bigg) \,,
\end{equation}
where $b$ is the harmonic oscillator length and $\Gamma$ is the gamma function. We now define
\begin{equation}
\gamma_{n,l,b}=\sqrt{\frac{2\Gamma(n+1)}{(b^2)^{l+\frac{3}{2}}\Gamma(n+l+\frac{3}{2})}}
\end{equation}
for simplicity. Taking the radial derivative and using (\ref{derLaguerre}), we have
\begin{equation}
\begin{split}
&\frac{dR_{n,l}}{dr}=\gamma_{n,l,b} \Big[ lr^{l-1} \exp\Big(-\frac{r^2}{2b^2}\Big)\,L_{n}^{l+\frac{1}{2}}\bigg(\frac{r^2}{b^2}\bigg) - \frac{r^{l+1}}{b^2} \exp\Big(-\frac{r^2}{2b^2}\Big)\,L_{n}^{l+\frac{1}{2}}\bigg(\frac{r^2}{b^2}\bigg) \\
&\qquad \qquad - \frac{2r^{l+1}}{b^2} \exp\Big(-\frac{r^2}{2b^2}\Big)\,L_{n-1}^{l+\frac{3}{2}}\bigg(\frac{r^2}{b^2}\bigg) \Big] \,.
\end{split}
\end{equation}
Now making use of (\ref{sumLaguerre}) and rewriting in terms of RHO functions,
\begin{equation}
\begin{split}
&\frac{dR_{n,l}}{dr} = \frac{l}{r}R_{n,l}(r) - \gamma_{n,l,b} \frac{r^{l+1}}{b^2} \exp\Big(-\frac{r^2}{2b^2}\Big) \Big[ L_{n}^{l+\frac{1}{2}}\bigg(\frac{r^2}{b^2}\bigg) + 2 L_{n-1}^{l+\frac{3}{2}}\bigg(\frac{r^2}{b^2}\bigg) \Big] \\
&\frac{dR_{n,l}}{dr} = \frac{l}{r}R_{n,l}(r) - \gamma_{n,l,b} \frac{r^{l+1}}{b^2} \exp\Big(-\frac{r^2}{2b^2}\Big) \Big[ L_{n}^{l+\frac{3}{2}}\bigg(\frac{r^2}{b^2}\bigg) + L_{n-1}^{l+\frac{3}{2}}\bigg(\frac{r^2}{b^2}\bigg) \Big] \\
&\frac{dR_{n,l}}{dr} = \frac{l}{r}R_{n,l}(r) - \frac{1}{b^2} \gamma_{n,l,b} \Bigg[ \frac{R_{n}^{l+1}(r)}{\gamma_{n,l+1,b}} +  \frac{R_{n-1}^{l+1}(r)}{\gamma_{n-1,l+1,b}}  \Bigg] \\
\end{split}
\end{equation}
we can derive our final result,
\begin{equation}
\begin{split}
&\frac{dR_{n,l}}{dr} = \frac{l}{r} R_{n,l} - \frac{1}{b} \bigg[ \sqrt{n+l+\frac{3}{2}} \, R_{n,l+1}(r) +\sqrt{n} \, R_{n-1,l+1}(r) \bigg]
\end{split}
\end{equation}
%

%%%% kinetic density del0 component
\subsection{Derivation of kinetic density}\label{sec_kdens}
We begin from the expression of the kinetic density in terms of the nonlocal density, Eq.~(\ref{kdens_eq}), and expand using the relation $\vec{\nabla} \cdot \vec{\nabla}\,' = \sum_{\mu=\pm1,0} (-1)^{\mu} \nabla_{\mu} \nabla_{-\mu}'$. We may write the kinetic density in component form as
\begin{equation}
\begin{split}
&{\tau}_\mathcal{N}(\vec{r}) = \bigg[ \nabla_{0} \nabla_{0}' {\rho}_\mathcal{N}(\vec{r},\vec{r}\,') - \nabla_{+1} \nabla_{-1}' {\rho}_\mathcal{N}(\vec{r},\vec{r}\,') - \nabla_{-1} \nabla_{+1}' {\rho}_\mathcal{N}(\vec{r},\vec{r}\,') \bigg]_{\vec{r}=\vec{r}\,'} \, . \\
\end{split}
\end{equation}
Let us consider solely the contribution of the $\nabla_{0} \nabla_{0}' {\rho}_\mathcal{N}(\vec{r},\vec{r}\,') \vert_{\vec{r}=\vec{r}\,'}$ component since the procedure is identical for all three components. We suppress the $ \mathcal{N} $ isospin label. It is convenient to rewrite this expression as follows,
\begin{equation}
\begin{split}
&\nabla_{0} \nabla_{0}' \rho(\vec{r},\vec{r}\,') \vert_{\vec{r}=\vec{r}\,'} = \sum_{n,l,n',l',K,k,m_{l},m_{l'}} \alpha_{n,l,n',l'}^{K,i,f}\, (l\,m_{l}\,l'\,m_{l'} \vert LM) \bigg[ \nabla_{0} R_{n,l}(r) Y_{l,m_{l}}^*(\hat{r}) \bigg] \bigg[ \nabla_{0}' R_{n',l'}(r') Y_{l',m_{l'}}^*(\hat{r}') \bigg] \, ,
\end{split}
\end{equation}
where $ \alpha_{n,l,n',l'}^{K,i,f} $ is defined in Eq.~(\ref{alphaeq}). Using the following relation from Ref. \cite{varshalovich1988quantum},
\begin{equation}\label{delRY}
\begin{split}
&\nabla_{0} R_{n,l}(r) Y_{l,m_{l}}^*(\hat{r}) = \sqrt{\frac{(l+1)^2-m_{l}^2}{(2l+1)(2l+3)}}\,\bigg(\frac{dR_{n,l}(r)}{dr} - \frac{l}{r}R_{n,l}(r)\bigg)\,Y_{l+1,m_{l}}^*(\hat{r}) \\
&\qquad + \sqrt{\frac{l^2-m_{l}^2}{(2l-1)(2l+1)}}\, \bigg(\frac{dR_{n,l}(r)}{dr} + \frac{l+1}{r}R_{n,l}(r)\bigg)\,Y_{l-1,m_{l}}^*(\hat{r}) \, , \\
\end{split}
\end{equation}
for both the $\nabla_{0}$ and $\nabla_{0}'$ terms, we then expand and evaluate our result at $\vec{r}=\vec{r}\,'$ to arrive at the formula shown in Eq.~(\ref{kdens_big}). Note that we group all spherical harmonics under the same collective index $L$ instead of having four separate angular momentum indices. We have
\begin{equation}\label{kdens_big}
\begin{split}
&\nabla_{0} \nabla_{0}' \rho(\vec{r},\vec{r}\,') \vert_{\vec{r}=\vec{r}\,'} \\
&\qquad = \sum_{n,l,n',l',K,k,m_{l},m_{l'}} \alpha_{n,l,n',l'}^{K}\, (l\,m_{l}\,l'\,m_{l'} \vert LM) \\
&\qquad \times \Bigg[ c_{00} \bigg(\frac{dR_{n,l}(r)}{dr} - \frac{l}{r}R_{n,l}(r)\bigg) \bigg(\frac{dR_{n',l'}(r)}{dr} - \frac{l'}{r}R_{n',l'}(r)\bigg) \\
&\qquad + c_{01} \bigg(\frac{dR_{n,l}(r)}{dr} - \frac{l}{r}R_{n,l}(r)\bigg) \bigg(\frac{dR_{n',l'}(r)}{dr} + \frac{l'+1}{r}R_{n',l'}(r)\bigg) \\
&\qquad + c_{02} \bigg(\frac{dR_{n,l}(r)}{dr} + \frac{l+1}{r}R_{n,l}(r)\bigg) \bigg(\frac{dR_{n',l'}(r)}{dr} - \frac{l'}{r}R_{n',l'}(r)\bigg) \\
&\qquad + c_{03} \bigg(\frac{dR_{n,l}(r)}{dr} + \frac{l+1}{r}R_{n,l}(r)\bigg) \bigg(\frac{dR_{n',l'}(r)}{dr} + \frac{l'+1}{r}R_{n',l'}(r)\bigg) \Bigg] \, Y_{LM}^*(\hat{r}) \,, \\
\end{split}
\end{equation}
where the $c_{0j}$ coefficients are complicated angular momentum factors. As an example, the $c_{00}$ factor is provided below in Eq.~(\ref{sample_c00}).
\begin{equation}\label{sample_c00}
\begin{split}
&c_{00}=\sqrt{\frac{(l+1)^2-m_{l}^2}{(2l+1)(2l+3)}}\sqrt{\frac{(l'+1)^2-m_{l'}^2}{(2l'+1)(2l'+3)}}\sqrt{\frac{(2l+3)(2l'+3)}{4\pi (2L+1)}} \\
&\qquad \times (l+1\,m_{l}\,l'+1\,m_{l'}\vert LM) \, (l+1\,0\,l'+1\,0\vert \, L0)
\end{split}
\end{equation}
A similar procedure can be performed for the $ {\nabla}_{+1} {\nabla}_{-1}' {\rho}_N $ and $ {\nabla}_{-1} {\nabla}_{+1}' {\rho}_N $ components of the kinetic density. The differences in the results are a result of the angular momentum coefficients, which have increments and decrements applied to the orbital angular momentum projection numbers as well.

\end{widetext}

%%%% acknowledgments section
\acknowledgments

We would like to acknowledge Angelo Calci for useful discussions and Witek Nazarewicz for helpful comments on this work. This work was supported by the NSERC Grant No. SAPIN-2016-00033. TRIUMF receives federal funding via a contribution agreement with the National Research Council of Canada. Computing support came from an INCITE Award on the Titan supercomputer of the Oak Ridge Leadership Computing Facility (OLCF) at ORNL, from Calcul Quebec, Westgrid and Compute Canada.

%%%% citations
%\bibliography{kdens}

\begin{thebibliography}{40}%
\makeatletter
\providecommand \@ifxundefined [1]{%
 \@ifx{#1\undefined}
}%
\providecommand \@ifnum [1]{%
 \ifnum #1\expandafter \@firstoftwo
 \else \expandafter \@secondoftwo
 \fi
}%
\providecommand \@ifx [1]{%
 \ifx #1\expandafter \@firstoftwo
 \else \expandafter \@secondoftwo
 \fi
}%
\providecommand \natexlab [1]{#1}%
\providecommand \enquote  [1]{``#1''}%
\providecommand \bibnamefont  [1]{#1}%
\providecommand \bibfnamefont [1]{#1}%
\providecommand \citenamefont [1]{#1}%
\providecommand \href@noop [0]{\@secondoftwo}%
\providecommand \href [0]{\begingroup \@sanitize@url \@href}%
\providecommand \@href[1]{\@@startlink{#1}\@@href}%
\providecommand \@@href[1]{\endgroup#1\@@endlink}%
\providecommand \@sanitize@url [0]{\catcode `\\12\catcode `\$12\catcode
  `\&12\catcode `\#12\catcode `\^12\catcode `\_12\catcode `\%12\relax}%
\providecommand \@@startlink[1]{}%
\providecommand \@@endlink[0]{}%
\providecommand \url  [0]{\begingroup\@sanitize@url \@url }%
\providecommand \@url [1]{\endgroup\@href {#1}{\urlprefix }}%
\providecommand \urlprefix  [0]{URL }%
\providecommand \Eprint [0]{\href }%
\providecommand \doibase [0]{http://dx.doi.org/}%
\providecommand \selectlanguage [0]{\@gobble}%
\providecommand \bibinfo  [0]{\@secondoftwo}%
\providecommand \bibfield  [0]{\@secondoftwo}%
\providecommand \translation [1]{[#1]}%
\providecommand \BibitemOpen [0]{}%
\providecommand \bibitemStop [0]{}%
\providecommand \bibitemNoStop [0]{.\EOS\space}%
\providecommand \EOS [0]{\spacefactor3000\relax}%
\providecommand \BibitemShut  [1]{\csname bibitem#1\endcsname}%
\let\auto@bib@innerbib\@empty
%</preamble>
\bibitem [{\citenamefont {Barrett}\ \emph {et~al.}(2013)\citenamefont
  {Barrett}, \citenamefont {Navr{\'a}til},\ and\ \citenamefont
  {Vary}}]{barrett2013ab}%
  \BibitemOpen
  \bibfield  {author} {\bibinfo {author} {\bibfnamefont {B.~R.}\ \bibnamefont
  {Barrett}}, \bibinfo {author} {\bibfnamefont {P.}~\bibnamefont
  {Navr{\'a}til}}, \ and\ \bibinfo {author} {\bibfnamefont {J.~P.}\
  \bibnamefont {Vary}},\ }\href@noop {} {\bibfield  {journal} {\bibinfo
  {journal} {Progress in Particle and Nuclear Physics}\ }\textbf {\bibinfo
  {volume} {69}},\ \bibinfo {pages} {131} (\bibinfo {year} {2013})}\BibitemShut
  {NoStop}%
\bibitem [{\citenamefont {Gennari}\ \emph {et~al.}(2018)\citenamefont
  {Gennari}, \citenamefont {Vorabbi}, \citenamefont {Calci},\ and\
  \citenamefont {Navr{\'a}til}}]{gennari2018microscopic}%
  \BibitemOpen
  \bibfield  {author} {\bibinfo {author} {\bibfnamefont {M.}~\bibnamefont
  {Gennari}}, \bibinfo {author} {\bibfnamefont {M.}~\bibnamefont {Vorabbi}},
  \bibinfo {author} {\bibfnamefont {A.}~\bibnamefont {Calci}}, \ and\ \bibinfo
  {author} {\bibfnamefont {P.}~\bibnamefont {Navr{\'a}til}},\ }\href@noop {}
  {\bibfield  {journal} {\bibinfo  {journal} {Physical Review C}\ }\textbf
  {\bibinfo {volume} {97}},\ \bibinfo {pages} {034619} (\bibinfo {year}
  {2018})}\BibitemShut {NoStop}%
\bibitem [{\citenamefont {Duguet}(2014)}]{duguet2014nuclear}%
  \BibitemOpen
  \bibfield  {author} {\bibinfo {author} {\bibfnamefont {T.}~\bibnamefont
  {Duguet}},\ }in\ \href@noop {} {\emph {\bibinfo {booktitle} {The Euroschool
  on Exotic Beams, Vol. IV}}}\ (\bibinfo  {publisher} {Springer},\ \bibinfo
  {year} {2014})\ pp.\ \bibinfo {pages} {293--350}\BibitemShut {NoStop}%
\bibitem [{\citenamefont {Vautherin}\ and\ \citenamefont
  {Brink}(1970)}]{vautherin1970hartree}%
  \BibitemOpen
  \bibfield  {author} {\bibinfo {author} {\bibfnamefont {D.}~\bibnamefont
  {Vautherin}}\ and\ \bibinfo {author} {\bibfnamefont {D.~M.}\ \bibnamefont
  {Brink}},\ }\href@noop {} {\bibfield  {journal} {\bibinfo  {journal} {Physics
  Letters B}\ }\textbf {\bibinfo {volume} {32}},\ \bibinfo {pages} {149}
  (\bibinfo {year} {1970})}\BibitemShut {NoStop}%
\bibitem [{\citenamefont {Vautherin}\ and\ \citenamefont
  {Brink}(1972)}]{vautherin1972hartree}%
  \BibitemOpen
  \bibfield  {author} {\bibinfo {author} {\bibfnamefont {D.}~\bibnamefont
  {Vautherin}}\ and\ \bibinfo {author} {\bibfnamefont {D.~M.}\ \bibnamefont
  {Brink}},\ }\href {\doibase 10.1103/PhysRevC.5.626} {\bibfield  {journal}
  {\bibinfo  {journal} {Physical Review C}\ }\textbf {\bibinfo {volume} {5}},\
  \bibinfo {pages} {626} (\bibinfo {year} {1972})}\BibitemShut {NoStop}%
\bibitem [{\citenamefont {Vautherin}(1973)}]{vautherin1973hartree}%
  \BibitemOpen
  \bibfield  {author} {\bibinfo {author} {\bibfnamefont {D.}~\bibnamefont
  {Vautherin}},\ }\href@noop {} {\bibfield  {journal} {\bibinfo  {journal}
  {Physical Review C}\ }\textbf {\bibinfo {volume} {7}},\ \bibinfo {pages}
  {296} (\bibinfo {year} {1973})}\BibitemShut {NoStop}%
\bibitem [{\citenamefont {Parr}\ and\ \citenamefont
  {Weitao}(1994)}]{parr1994density}%
  \BibitemOpen
  \bibfield  {author} {\bibinfo {author} {\bibfnamefont {R.}~\bibnamefont
  {Parr}}\ and\ \bibinfo {author} {\bibfnamefont {Y.}~\bibnamefont {Weitao}},\
  }\href {https://books.google.ca/books?id=mGOpScSIwU4C} {\emph {\bibinfo
  {title} {Density-Functional Theory of Atoms and Molecules}}},\ International
  Series of Monographs on Chemistry\ (\bibinfo  {publisher} {Oxford University
  Press},\ \bibinfo {year} {1994})\BibitemShut {NoStop}%
\bibitem [{\citenamefont {Dreizler}(2013)}]{gross2013density}%
  \BibitemOpen
  \bibfield  {author} {\bibinfo {author} {\bibfnamefont {R.~M.}\ \bibnamefont
  {Dreizler}},\ }\href@noop {} {\emph {\bibinfo {title} {Density Functional
  Theory}}}\ (\bibinfo  {publisher} {Springer Berlin Heidelberg},\ \bibinfo
  {year} {2013})\BibitemShut {NoStop}%
\bibitem [{\citenamefont {Bhattacharyya}\ and\ \citenamefont
  {Furnstahl}(2005)}]{bhattacharyya2005kinetic}%
  \BibitemOpen
  \bibfield  {author} {\bibinfo {author} {\bibfnamefont {A.}~\bibnamefont
  {Bhattacharyya}}\ and\ \bibinfo {author} {\bibfnamefont {R.~J.}~\bibnamefont
  {Furnstahl}},\ }\href@noop {} {\bibfield  {journal} {\bibinfo  {journal}
  {Nuclear Physics A}\ }\textbf {\bibinfo {volume} {747}},\ \bibinfo {pages}
  {268} (\bibinfo {year} {2005})}\BibitemShut {NoStop}%
\bibitem [{\citenamefont {Hofmann}\ and\ \citenamefont
  {Lenske}(1998)}]{hofmann1998hartree}%
  \BibitemOpen
  \bibfield  {author} {\bibinfo {author} {\bibfnamefont {F.}~\bibnamefont
  {Hofmann}}\ and\ \bibinfo {author} {\bibfnamefont {H.}~\bibnamefont
  {Lenske}},\ }\href@noop {} {\bibfield  {journal} {\bibinfo  {journal}
  {Physical Review C}\ }\textbf {\bibinfo {volume} {57}},\ \bibinfo {pages}
  {2281} (\bibinfo {year} {1998})}\BibitemShut {NoStop}%
\bibitem [{\citenamefont {Erler}\ \emph {et~al.}(2012)\citenamefont {Erler},
  \citenamefont {Birge}, \citenamefont {Kortelainen}, \citenamefont
  {Nazarewicz}, \citenamefont {Olsen}, \citenamefont {Perhac},\ and\
  \citenamefont {Stoitsov}}]{erler2012limits}%
  \BibitemOpen
  \bibfield  {author} {\bibinfo {author} {\bibfnamefont {J.}~\bibnamefont
  {Erler}}, \bibinfo {author} {\bibfnamefont {N.}~\bibnamefont {Birge}},
  \bibinfo {author} {\bibfnamefont {M.}~\bibnamefont {Kortelainen}}, \bibinfo
  {author} {\bibfnamefont {W.}~\bibnamefont {Nazarewicz}}, \bibinfo {author}
  {\bibfnamefont {E.}~\bibnamefont {Olsen}}, \bibinfo {author} {\bibfnamefont
  {A.~M.}\ \bibnamefont {Perhac}}, \ and\ \bibinfo {author} {\bibfnamefont
  {M.~V.}~\bibnamefont {Stoitsov}},\ }\href@noop {} {\bibfield  {journal}
  {\bibinfo  {journal} {Nature}\ }\textbf {\bibinfo {volume} {486}},\ \bibinfo
  {pages} {509} (\bibinfo {year} {2012})}\BibitemShut {NoStop}%
\bibitem [{\citenamefont {Kortelainen}\ \emph {et~al.}(2010)\citenamefont
  {Kortelainen}, \citenamefont {Lesinski}, \citenamefont {Mor{\'e}},
  \citenamefont {Nazarewicz}, \citenamefont {Sarich}, \citenamefont {Schunck},
  \citenamefont {Stoitsov},\ and\ \citenamefont
  {Wild}}]{kortelainen2010nuclear}%
  \BibitemOpen
  \bibfield  {author} {\bibinfo {author} {\bibfnamefont {M.}~\bibnamefont
  {Kortelainen}}, \bibinfo {author} {\bibfnamefont {T.}~\bibnamefont
  {Lesinski}}, \bibinfo {author} {\bibfnamefont {J.}~\bibnamefont {Mor{\'e}}},
  \bibinfo {author} {\bibfnamefont {W.}~\bibnamefont {Nazarewicz}}, \bibinfo
  {author} {\bibfnamefont {J.}~\bibnamefont {Sarich}}, \bibinfo {author}
  {\bibfnamefont {N.}~\bibnamefont {Schunck}}, \bibinfo {author} {\bibfnamefont
  {M.~V.}~\bibnamefont {Stoitsov}}, \ and\ \bibinfo {author} {\bibfnamefont
  {S.~M.}~\bibnamefont {Wild}},\ }\href@noop {} {\bibfield  {journal} {\bibinfo
  {journal} {Physical Review C}\ }\textbf {\bibinfo {volume} {82}},\ \bibinfo
  {pages} {024313} (\bibinfo {year} {2010})}\BibitemShut {NoStop}%
\bibitem [{\citenamefont {Kortelainen}\ \emph {et~al.}(2012)\citenamefont
  {Kortelainen}, \citenamefont {McDonnell}, \citenamefont {Nazarewicz},
  \citenamefont {Reinhard}, \citenamefont {Sarich}, \citenamefont {Schunck},
  \citenamefont {Stoitsov},\ and\ \citenamefont
  {Wild}}]{kortelainen2012nuclear}%
  \BibitemOpen
  \bibfield  {author} {\bibinfo {author} {\bibfnamefont {M.}~\bibnamefont
  {Kortelainen}}, \bibinfo {author} {\bibfnamefont {J.}~\bibnamefont
  {McDonnell}}, \bibinfo {author} {\bibfnamefont {W.}~\bibnamefont
  {Nazarewicz}}, \bibinfo {author} {\bibfnamefont {P.-G.}\ \bibnamefont
  {Reinhard}}, \bibinfo {author} {\bibfnamefont {J.}~\bibnamefont {Sarich}},
  \bibinfo {author} {\bibfnamefont {N.}~\bibnamefont {Schunck}}, \bibinfo
  {author} {\bibfnamefont {M.~V.}~\bibnamefont {Stoitsov}}, \ and\ \bibinfo
  {author} {\bibfnamefont {S.~M.}~\bibnamefont {Wild}},\ }\href@noop {} {\bibfield
   {journal} {\bibinfo  {journal} {Physical Review C}\ }\textbf {\bibinfo
  {volume} {85}},\ \bibinfo {pages} {024304} (\bibinfo {year}
  {2012})}\BibitemShut {NoStop}%
\bibitem [{\citenamefont {Kortelainen}\ \emph {et~al.}(2014)\citenamefont
  {Kortelainen}, \citenamefont {McDonnell}, \citenamefont {Nazarewicz},
  \citenamefont {Olsen}, \citenamefont {Reinhard}, \citenamefont {Sarich},
  \citenamefont {Schunck}, \citenamefont {Wild}, \citenamefont {Davesne},
  \citenamefont {Erler} \emph {et~al.}}]{kortelainen2014nuclear}%
  \BibitemOpen
  \bibfield  {author} {\bibinfo {author} {\bibfnamefont {M.}~\bibnamefont
  {Kortelainen}}, \bibinfo {author} {\bibfnamefont {J.}~\bibnamefont
  {McDonnell}}, \bibinfo {author} {\bibfnamefont {W.}~\bibnamefont
  {Nazarewicz}}, \bibinfo {author} {\bibfnamefont {E.}~\bibnamefont {Olsen}},
  \bibinfo {author} {\bibfnamefont {P.-G.}\ \bibnamefont {Reinhard}}, \bibinfo
  {author} {\bibfnamefont {J.}~\bibnamefont {Sarich}}, \bibinfo {author}
  {\bibfnamefont {N.}~\bibnamefont {Schunck}}, \bibinfo {author} {\bibfnamefont
  {S.~M.}~\bibnamefont {Wild}}, \bibinfo {author} {\bibfnamefont {D.}~\bibnamefont
  {Davesne}}, \bibinfo {author} {\bibfnamefont {J.}~\bibnamefont {Erler}},
  \emph {et~al.},\ }\href@noop {} {\bibfield  {journal} {\bibinfo  {journal}
  {Physical Review C}\ }\textbf {\bibinfo {volume} {89}},\ \bibinfo {pages}
  {054314} (\bibinfo {year} {2014})}\BibitemShut {NoStop}%
\bibitem [{\citenamefont {Navarro~P\'erez}\ \emph {et~al.}(2018)\citenamefont
  {Navarro~P\'erez}, \citenamefont {Schunck}, \citenamefont {Dyhdalo},
  \citenamefont {Furnstahl},\ and\ \citenamefont
  {Bogner}}]{microscopicEDF2018}%
  \BibitemOpen
  \bibfield  {author} {\bibinfo {author} {\bibfnamefont {R.}~\bibnamefont
  {Navarro~P\'erez}}, \bibinfo {author} {\bibfnamefont {N.}~\bibnamefont
  {Schunck}}, \bibinfo {author} {\bibfnamefont {A.}~\bibnamefont {Dyhdalo}},
  \bibinfo {author} {\bibfnamefont {R.~J.}\ \bibnamefont {Furnstahl}}, \ and\
  \bibinfo {author} {\bibfnamefont {S.~K.}\ \bibnamefont {Bogner}},\ }\href
  {\doibase 10.1103/PhysRevC.97.054304} {\bibfield  {journal} {\bibinfo
  {journal} {Phys. Rev. C}\ }\textbf {\bibinfo {volume} {97}},\ \bibinfo
  {pages} {054304} (\bibinfo {year} {2018})}\BibitemShut {NoStop}%
\bibitem [{\citenamefont {Lacroix}\ \emph {et~al.}(2017)\citenamefont
  {Lacroix}, \citenamefont {Boulet}, \citenamefont {Grasso},\ and\
  \citenamefont {Yang}}]{lacroix2017bare}%
  \BibitemOpen
  \bibfield  {author} {\bibinfo {author} {\bibfnamefont {D.}~\bibnamefont
  {Lacroix}}, \bibinfo {author} {\bibfnamefont {A.}~\bibnamefont {Boulet}},
  \bibinfo {author} {\bibfnamefont {M.}~\bibnamefont {Grasso}}, \ and\ \bibinfo
  {author} {\bibfnamefont {C.-J.}\ \bibnamefont {Yang}},\ }\href@noop {}
  {\bibfield  {journal} {\bibinfo  {journal} {Physical Review C}\ }\textbf
  {\bibinfo {volume} {95}},\ \bibinfo {pages} {054306} (\bibinfo {year}
  {2017})}\BibitemShut {NoStop}%
\bibitem [{\citenamefont {Grasso}(2018)}]{grasso2018effective}%
  \BibitemOpen
  \bibfield  {author} {\bibinfo {author} {\bibfnamefont {M.}~\bibnamefont
  {Grasso}},\ }\href@noop {} {\bibfield  {journal} {\bibinfo  {journal} {arXiv
  preprint arXiv:1811.01039}\ } (\bibinfo {year} {2018})}\BibitemShut {NoStop}%
\bibitem [{\citenamefont {Bonnard}\ \emph {et~al.}(2018)\citenamefont
  {Bonnard}, \citenamefont {Grasso},\ and\ \citenamefont
  {Lacroix}}]{bonnard2018effective}%
  \BibitemOpen
  \bibfield  {author} {\bibinfo {author} {\bibfnamefont {J.}~\bibnamefont
  {Bonnard}}, \bibinfo {author} {\bibfnamefont {M.}~\bibnamefont {Grasso}}, \
  and\ \bibinfo {author} {\bibfnamefont {D.}~\bibnamefont {Lacroix}},\ }\href
  {\doibase 10.1103/PhysRevC.98.034319} {\bibfield  {journal} {\bibinfo
  {journal} {Physical Review C}\ }\textbf {\bibinfo {volume} {98}},\ \bibinfo
  {pages} {034319} (\bibinfo {year} {2018})}\BibitemShut {NoStop}%
\bibitem [{\citenamefont {Wegner}(1994)}]{wegner1994flow}%
  \BibitemOpen
  \bibfield  {author} {\bibinfo {author} {\bibfnamefont {F.}~\bibnamefont
  {Wegner}},\ }\href@noop {} {\bibfield  {journal} {\bibinfo  {journal}
  {Annalen der physik}\ }\textbf {\bibinfo {volume} {506}},\ \bibinfo {pages}
  {77} (\bibinfo {year} {1994})}\BibitemShut {NoStop}%
\bibitem [{\citenamefont {Bogner}\ \emph {et~al.}(2007)\citenamefont {Bogner},
  \citenamefont {Furnstahl},\ and\ \citenamefont
  {Perry}}]{bogner2007similarity}%
  \BibitemOpen
  \bibfield  {author} {\bibinfo {author} {\bibfnamefont {S.~K.}~\bibnamefont
  {Bogner}}, \bibinfo {author} {\bibfnamefont {R.~J.}~\bibnamefont {Furnstahl}}, \
  and\ \bibinfo {author} {\bibfnamefont {R.~J.}~\bibnamefont {Perry}},\
  }\href@noop {} {\bibfield  {journal} {\bibinfo  {journal} {Physical Review
  C}\ }\textbf {\bibinfo {volume} {75}},\ \bibinfo {pages} {061001} (\bibinfo
  {year} {2007})}\BibitemShut {NoStop}%
\bibitem [{\citenamefont {Roth}\ \emph {et~al.}(2008)\citenamefont {Roth},
  \citenamefont {Reinhardt},\ and\ \citenamefont {Hergert}}]{roth2008unitary}%
  \BibitemOpen
  \bibfield  {author} {\bibinfo {author} {\bibfnamefont {R.}~\bibnamefont
  {Roth}}, \bibinfo {author} {\bibfnamefont {S.}~\bibnamefont {Reinhardt}}, \
  and\ \bibinfo {author} {\bibfnamefont {H.}~\bibnamefont {Hergert}},\
  }\href@noop {} {\bibfield  {journal} {\bibinfo  {journal} {Physical Review
  C}\ }\textbf {\bibinfo {volume} {77}},\ \bibinfo {pages} {064003} (\bibinfo
  {year} {2008})}\BibitemShut {NoStop}%
\bibitem [{\citenamefont {Bogner}\ \emph {et~al.}(2010)\citenamefont {Bogner},
  \citenamefont {Furnstahl},\ and\ \citenamefont {Schwenk}}]{bogner2010low}%
  \BibitemOpen
  \bibfield  {author} {\bibinfo {author} {\bibfnamefont {S.~K.}~\bibnamefont
  {Bogner}}, \bibinfo {author} {\bibfnamefont {R.~J.}~\bibnamefont {Furnstahl}}, \
  and\ \bibinfo {author} {\bibfnamefont {A.}~\bibnamefont {Schwenk}},\
  }\href@noop {} {\bibfield  {journal} {\bibinfo  {journal} {Progress in
  Particle and Nuclear Physics}\ }\textbf {\bibinfo {volume} {65}},\ \bibinfo
  {pages} {94} (\bibinfo {year} {2010})}\BibitemShut {NoStop}%
\bibitem [{\citenamefont {Jurgenson}\ \emph {et~al.}(2009)\citenamefont
  {Jurgenson}, \citenamefont {Navratil},\ and\ \citenamefont
  {Furnstahl}}]{jurgenson2009evolution}%
  \BibitemOpen
  \bibfield  {author} {\bibinfo {author} {\bibfnamefont {E.~D.}~\bibnamefont
  {Jurgenson}}, \bibinfo {author} {\bibfnamefont {P.}~\bibnamefont {Navratil}},
  \ and\ \bibinfo {author} {\bibfnamefont {R.~J.}~\bibnamefont {Furnstahl}},\
  }\href@noop {} {\bibfield  {journal} {\bibinfo  {journal} {Physical review
  letters}\ }\textbf {\bibinfo {volume} {103}},\ \bibinfo {pages} {082501}
  (\bibinfo {year} {2009})}\BibitemShut {NoStop}%
\bibitem [{\citenamefont {Entem}\ \emph {et~al.}(2015)\citenamefont {Entem},
  \citenamefont {Kaiser}, \citenamefont {Machleidt},\ and\ \citenamefont
  {Nosyk}}]{entem2015peripheral}%
  \BibitemOpen
  \bibfield  {author} {\bibinfo {author} {\bibfnamefont {D.~R.}~\bibnamefont
  {Entem}}, \bibinfo {author} {\bibfnamefont {N.}~\bibnamefont {Kaiser}},
  \bibinfo {author} {\bibfnamefont {R.}~\bibnamefont {Machleidt}}, \ and\
  \bibinfo {author} {\bibfnamefont {Y.}~\bibnamefont {Nosyk}},\ }\href@noop {}
  {\bibfield  {journal} {\bibinfo  {journal} {Physical Review C}\ }\textbf
  {\bibinfo {volume} {91}},\ \bibinfo {pages} {014002} (\bibinfo {year}
  {2015})}\BibitemShut {NoStop}%
\bibitem [{\citenamefont {Entem}\ \emph {et~al.}(2017)\citenamefont {Entem},
  \citenamefont {Machleidt},\ and\ \citenamefont {Nosyk}}]{entem2017high}%
  \BibitemOpen
  \bibfield  {author} {\bibinfo {author} {\bibfnamefont {D.~R.}\ \bibnamefont
  {Entem}}, \bibinfo {author} {\bibfnamefont {R.}~\bibnamefont {Machleidt}}, \
  and\ \bibinfo {author} {\bibfnamefont {Y.}~\bibnamefont {Nosyk}},\ }\href
  {\doibase 10.1103/PhysRevC.96.024004} {\bibfield  {journal} {\bibinfo
  {journal} {Phys. Rev. C}\ }\textbf {\bibinfo {volume} {96}},\ \bibinfo
  {pages} {024004} (\bibinfo {year} {2017})}\BibitemShut {NoStop}%
\bibitem [{\citenamefont {Navratil}(2007)}]{navratil2007local}%
  \BibitemOpen
  \bibfield  {author} {\bibinfo {author} {\bibfnamefont {P.}~\bibnamefont
  {Navratil}},\ }\href@noop {} {\bibfield  {journal} {\bibinfo  {journal}
  {Few-Body Systems}\ }\textbf {\bibinfo {volume} {41}},\ \bibinfo {pages}
  {117} (\bibinfo {year} {2007})}\BibitemShut {NoStop}%
\bibitem [{\citenamefont {Roth}\ and\ \citenamefont
  {Navr{\'a}til}(2007)}]{roth2007ab}%
  \BibitemOpen
  \bibfield  {author} {\bibinfo {author} {\bibfnamefont {R.}~\bibnamefont
  {Roth}}\ and\ \bibinfo {author} {\bibfnamefont {P.}~\bibnamefont
  {Navr{\'a}til}},\ }\href@noop {} {\bibfield  {journal} {\bibinfo  {journal}
  {Physical review letters}\ }\textbf {\bibinfo {volume} {99}},\ \bibinfo
  {pages} {092501} (\bibinfo {year} {2007})}\BibitemShut {NoStop}%
\bibitem [{\citenamefont {Roth}(2009)}]{roth2009importance}%
  \BibitemOpen
  \bibfield  {author} {\bibinfo {author} {\bibfnamefont {R.}~\bibnamefont
  {Roth}},\ }\href@noop {} {\bibfield  {journal} {\bibinfo  {journal} {Physical
  Review C}\ }\textbf {\bibinfo {volume} {79}},\ \bibinfo {pages} {064324}
  (\bibinfo {year} {2009})}\BibitemShut {NoStop}%
\bibitem [{\citenamefont {Navr{\'a}til}(2004)}]{navratil2004translationally}%
  \BibitemOpen
  \bibfield  {author} {\bibinfo {author} {\bibfnamefont {P.}~\bibnamefont
  {Navr{\'a}til}},\ }\href@noop {} {\bibfield  {journal} {\bibinfo  {journal}
  {Physical Review C}\ }\textbf {\bibinfo {volume} {70}},\ \bibinfo {pages}
  {014317} (\bibinfo {year} {2004})}\BibitemShut {NoStop}%
\bibitem [{\citenamefont {Burrows}\ \emph {et~al.}(2018)\citenamefont
  {Burrows}, \citenamefont {Elster}, \citenamefont {Popa}, \citenamefont
  {Launey}, \citenamefont {Nogga},\ and\ \citenamefont
  {Maris}}]{burrows2018ab}%
  \BibitemOpen
  \bibfield  {author} {\bibinfo {author} {\bibfnamefont {M.}~\bibnamefont
  {Burrows}}, \bibinfo {author} {\bibfnamefont {C.}~\bibnamefont {Elster}},
  \bibinfo {author} {\bibfnamefont {G.}~\bibnamefont {Popa}}, \bibinfo {author}
  {\bibfnamefont {K.~D.}~\bibnamefont {Launey}}, \bibinfo {author} {\bibfnamefont
  {A.}~\bibnamefont {Nogga}}, \ and\ \bibinfo {author} {\bibfnamefont
  {P.}~\bibnamefont {Maris}},\ }\href@noop {} {\bibfield  {journal} {\bibinfo
  {journal} {Physical Review C}\ }\textbf {\bibinfo {volume} {97}},\ \bibinfo
  {pages} {024325} (\bibinfo {year} {2018})}\BibitemShut {NoStop}%
\bibitem [{\citenamefont {Giraud}(2008)}]{giraud2008labframe}%
  \BibitemOpen
  \bibfield  {author} {\bibinfo {author} {\bibfnamefont {B.~G.}\ \bibnamefont
  {Giraud}},\ }\href {\doibase 10.1103/PhysRevC.77.014311} {\bibfield
  {journal} {\bibinfo  {journal} {Physical Review C}\ }\textbf {\bibinfo
  {volume} {77}},\ \bibinfo {pages} {014311} (\bibinfo {year}
  {2008})}\BibitemShut {NoStop}%
\bibitem [{\citenamefont {Navr{\'a}til}\ \emph {et~al.}(2000)\citenamefont
  {Navr{\'a}til}, \citenamefont {Kamuntavi{\v{c}}ius},\ and\ \citenamefont
  {Barrett}}]{navratil2000few}%
  \BibitemOpen
  \bibfield  {author} {\bibinfo {author} {\bibfnamefont {P.}~\bibnamefont
  {Navr{\'a}til}}, \bibinfo {author} {\bibfnamefont {G.~P.}~\bibnamefont
  {Kamuntavi{\v{c}}ius}}, \ and\ \bibinfo {author} {\bibfnamefont
  {B.~R.}~\bibnamefont {Barrett}},\ }\href@noop {} {\bibfield  {journal} {\bibinfo
   {journal} {Physical Review C}\ }\textbf {\bibinfo {volume} {61}},\ \bibinfo
  {pages} {044001} (\bibinfo {year} {2000})}\BibitemShut {NoStop}%
\bibitem [{\citenamefont {Trlifaj}(1972)}]{trlifaj1972simple}%
  \BibitemOpen
  \bibfield  {author} {\bibinfo {author} {\bibfnamefont {L.}~\bibnamefont
  {Trlifaj}},\ }\href@noop {} {\bibfield  {journal} {\bibinfo  {journal}
  {Physical Review C}\ }\textbf {\bibinfo {volume} {5}},\ \bibinfo {pages}
  {1534} (\bibinfo {year} {1972})}\BibitemShut {NoStop}%
\bibitem [{\citenamefont {Borycki}\ \emph {et~al.}(2006)\citenamefont
  {Borycki}, \citenamefont {Dobaczewski}, \citenamefont {Nazarewicz},\ and\
  \citenamefont {Stoitsov}}]{borycki2006pairing}%
  \BibitemOpen
  \bibfield  {author} {\bibinfo {author} {\bibfnamefont {P.~J.}~\bibnamefont
  {Borycki}}, \bibinfo {author} {\bibfnamefont {J.}~\bibnamefont
  {Dobaczewski}}, \bibinfo {author} {\bibfnamefont {W.}~\bibnamefont
  {Nazarewicz}}, \ and\ \bibinfo {author} {\bibfnamefont {M.~V.}~\bibnamefont
  {Stoitsov}},\ }\href@noop {} {\bibfield  {journal} {\bibinfo  {journal}
  {Physical Review C}\ }\textbf {\bibinfo {volume} {73}},\ \bibinfo {pages}
  {044319} (\bibinfo {year} {2006})}\BibitemShut {NoStop}%
\bibitem [{\citenamefont {Dobaczewski}\ and\ \citenamefont
  {Dudek}(1995)}]{dobaczewski1995time}%
  \BibitemOpen
  \bibfield  {author} {\bibinfo {author} {\bibfnamefont {J.}~\bibnamefont
  {Dobaczewski}}\ and\ \bibinfo {author} {\bibfnamefont {J.}~\bibnamefont
  {Dudek}},\ }\href@noop {} {\bibfield  {journal} {\bibinfo  {journal}
  {Physical Review C}\ }\textbf {\bibinfo {volume} {52}},\ \bibinfo {pages}
  {1827} (\bibinfo {year} {1995})}\BibitemShut {NoStop}%
\bibitem [{\citenamefont {Engel}\ \emph {et~al.}(1975)\citenamefont {Engel},
  \citenamefont {Brink}, \citenamefont {Goeke}, \citenamefont {Krieger},\ and\
  \citenamefont {Vautherin}}]{engel1975time}%
  \BibitemOpen
  \bibfield  {author} {\bibinfo {author} {\bibfnamefont {Y.}~\bibnamefont
  {Engel}}, \bibinfo {author} {\bibfnamefont {D.}~\bibnamefont {Brink}},
  \bibinfo {author} {\bibfnamefont {K.}~\bibnamefont {Goeke}}, \bibinfo
  {author} {\bibfnamefont {S.}~\bibnamefont {Krieger}}, \ and\ \bibinfo
  {author} {\bibfnamefont {D.}~\bibnamefont {Vautherin}},\ }\href@noop {}
  {\bibfield  {journal} {\bibinfo  {journal} {Nuclear Physics A}\ }\textbf
  {\bibinfo {volume} {249}},\ \bibinfo {pages} {215} (\bibinfo {year}
  {1975})}\BibitemShut {NoStop}%
\bibitem [{\citenamefont {Varshalovich}\ \emph {et~al.}(1988)\citenamefont
  {Varshalovich}, \citenamefont {Moskalev},\ and\ \citenamefont
  {Khersonskii}}]{varshalovich1988quantum}%
  \BibitemOpen
  \bibfield  {author} {\bibinfo {author} {\bibfnamefont {D.~A.}\ \bibnamefont
  {Varshalovich}}, \bibinfo {author} {\bibfnamefont {A.~N.}\ \bibnamefont
  {Moskalev}}, \ and\ \bibinfo {author} {\bibfnamefont {V.~K.}\ \bibnamefont
  {Khersonskii}},\ }\href@noop {} {\emph {\bibinfo {title} {Quantum theory of
  angular momentum}}}\ (\bibinfo  {publisher} {World Scientific},\ \bibinfo
  {year} {1988})\BibitemShut {NoStop}%
\bibitem [{\citenamefont {Dobaczewski}\ \emph {et~al.}(2003)\citenamefont
  {Dobaczewski}, \citenamefont {Nazarewicz},\ and\ \citenamefont
  {Stoitsov}}]{dobaczewski2003nuclear}%
  \BibitemOpen
  \bibfield  {author} {\bibinfo {author} {\bibfnamefont {J.}~\bibnamefont
  {Dobaczewski}}, \bibinfo {author} {\bibfnamefont {W.}~\bibnamefont
  {Nazarewicz}}, \ and\ \bibinfo {author} {\bibfnamefont {M.~V.}~\bibnamefont
  {Stoitsov}},\ }in\ \href@noop {} {\emph {\bibinfo {booktitle} {Exotic Nuclei
  and Atomic Masses}}}\ (\bibinfo  {publisher} {Springer},\ \bibinfo {year}
  {2003})\ pp.\ \bibinfo {pages} {55--60}\BibitemShut {NoStop}%
\bibitem [{\citenamefont {Bender}\ \emph {et~al.}(2003)\citenamefont {Bender},
  \citenamefont {Heenen},\ and\ \citenamefont {Reinhard}}]{bender2003self}%
  \BibitemOpen
  \bibfield  {author} {\bibinfo {author} {\bibfnamefont {M.}~\bibnamefont
  {Bender}}, \bibinfo {author} {\bibfnamefont {P.-H.}\ \bibnamefont {Heenen}},
  \ and\ \bibinfo {author} {\bibfnamefont {P.-G.}\ \bibnamefont {Reinhard}},\
  }\href@noop {} {\bibfield  {journal} {\bibinfo  {journal} {Reviews of Modern
  Physics}\ }\textbf {\bibinfo {volume} {75}},\ \bibinfo {pages} {121}
  (\bibinfo {year} {2003})}\BibitemShut {NoStop}%
\bibitem [{\citenamefont {Bender}\ \emph {et~al.}(2000)\citenamefont {Bender},
  \citenamefont {Rutz}, \citenamefont {Reinhard},\ and\ \citenamefont
  {Maruhn}}]{bender2000consequences}%
  \BibitemOpen
  \bibfield  {author} {\bibinfo {author} {\bibfnamefont {M.}~\bibnamefont
  {Bender}}, \bibinfo {author} {\bibfnamefont {K.}~\bibnamefont {Rutz}},
  \bibinfo {author} {\bibfnamefont {P.-G.}\ \bibnamefont {Reinhard}}, \ and\
  \bibinfo {author} {\bibfnamefont {J.}~\bibnamefont {Maruhn}},\ }\href@noop {}
  {\bibfield  {journal} {\bibinfo  {journal} {The European Physical Journal
  A-Hadrons and Nuclei}\ }\textbf {\bibinfo {volume} {7}},\ \bibinfo {pages}
  {467} (\bibinfo {year} {2000})}\BibitemShut {NoStop}%
\end{thebibliography}

%merlin.mbs apsrev4-1.bst 2010-07-25 4.21a (PWD, AO, DPC) hacked
%Control: key (0)
%Control: author (8) initials jnrlst
%Control: editor formatted (1) identically to author
%Control: production of article title (-1) disabled
%Control: page (0) single
%Control: year (1) truncated
%Control: production of eprint (0) enabled
%

\end{document}